\documentclass[10pt,oneside]{article}                   

\usepackage{graphicx}

\usepackage{amsmath,amssymb,color,graphicx,pdfsync,overpic,color,epstopdf,rotating,dashrule,float,bm,multicol,setspace,enumerate}
\usepackage[affil-it]{authblk}

\usepackage{amsthm}
\usepackage{mathrsfs}
\usepackage{mathtools}
\usepackage{graphicx}
\usepackage{textcomp}
\usepackage{setspace}
\usepackage{listings}
\usepackage{fullpage}
\usepackage{natbib}
\usepackage{array}
\usepackage{comment}
\usepackage{tikz}
\usetikzlibrary{calc,arrows,automata,topaths}

\newcommand\Rey{\mbox{\textit{Re}}}  

\newsavebox{\astrutbox}
\sbox{\astrutbox}{\rule[-5pt]{0pt}{20pt}}

\newcommand{\defeq}{\vcentcolon=}

\newcommand{\bx}{{ \bf x}}

\newcommand{\bn}{{ \bf n}}

\newcommand{\bz}{{ \bf z}}

\newcommand{\bE}{\ensuremath{\mathbb{E}}}

\def\ip<#1,#2>{\left\langle #1,#2\right\rangle}
\newtheorem{algorithm}{Algorithm}

\newcommand*{\todo}[1]{}

\graphicspath{{SubmissionVersion/}{Figures/}}

\usepackage{hyperref}			
\usepackage[all]{hypcap}		
\hypersetup{colorlinks=true,	
	linkcolor=red,			    
	citecolor=blue,		        
   urlcolor=cyan               
	}
	
\title{Characterizing and correcting for the effect of sensor noise in the dynamic mode decomposition
}

\author{Scott T. M. Dawson\thanks{\href{mailto:stdawson@princeton.edu}{stdawson@princeton.edu}; Corresponding author},      
	   \  Maziar S. Hemati,     
	 \    Matthew O. Williams  \ \&
      \    Clarence W. Rowley 
}

\begin{document}

\maketitle

\begin{abstract}

Dynamic mode decomposition (DMD) provides a practical means of extracting insightful dynamical information from fluids datasets. 
Like any data processing technique, DMD's usefulness is limited by its ability to extract real and accurate dynamical features from noise-corrupted data. 
Here we show analytically that DMD is biased to sensor noise, and quantify how this bias depends on the size and noise level of the data.
We present three modifications to DMD that can be used to remove this bias: (i) a direct correction of the identified bias using known noise properties, (ii) combining the results of performing DMD forwards and backwards in time, and (iii) a total least-squares-inspired algorithm. 
We discuss the relative merits of each algorithm, and demonstrate the performance of these modifications on a range of synthetic, numerical, and experimental datasets. 
We further compare our modified DMD algorithms with other variants proposed in recent literature.

\end{abstract}

\section{Introduction}
\label{sec:intro}

With advances in both experimental techniques and equipment, and computational power and storage capacity, researchers in fluid dynamics can now generate more high-fidelity data than ever before. 
The presence of increasingly large data sets calls for appropriate data analysis techniques, that are able to extract tractable and physically relevant information from the data.  
Dynamic mode decomposition allows for the identification and analysis of dynamical features of time-evolving fluid flows, using data obtained from either experiments or simulations. 
In contrast to other data-driven modal decompositions such as the proper orthogonal decomposition (POD), DMD allows for spatial modes to be identified that can be directly associated with characteristic frequencies and growth/decay rates. 
Following its conception, DMD was quickly shown to be useful in extracting
dynamical features in both experimental and numerical data
\citep{schmid2008,schmid2010DMD}. It has subsequently been used to gain dynamic
insight on a wide range of problems arising in fluid mechanics
\citep[e.g.,][]{schmid2011tcfd,schmid2011expfluids,jardin2012dmd} and other fields
\citep[e.g.,][]{grosek2014dmd}.

DMD has a strong connection to Koopman operator theory \citep{koopman1931hamiltonian,mezic2005spectral}, as exposed in \cite{rowley2009spectral}, and further reviewed in \cite{mezic2013koopman}, which can justify its use in analyzing nonlinear dynamical systems. Since its original formulation, numerous modifications and extensions have been made to DMD. \cite{chen2011variants} highlights the connection that DMD shares with traditional Fourier analysis, as well as proposing an optimized algorithm that recasts DMD as an optimal dimensionality reduction problem. 
This concept of finding only the dynamically important modes has also been considered in subsequent works of \cite{wynn2013omd} and \cite{jovanovic2014dmdsp}.  All of these works are motivated, in part, by the fact that by default DMD will output as many modes as there are pairs of snapshots (assuming that the length of the snapshot vector is greater than the number of snapshots), which is arbitrary with respect to the dynamical system under consideration. In reality, one would prefer to output only the modes and eigenvalues that are present (or dominant) in the data. When the data is corrupted by noise (as will always be the case to some degree, especially for experimental data), this process becomes nontrivial, since noisy data might have a numerical rank far larger than the dimension of the governing dynamics of the system. Further to this, one cannot expect to have a clean partition into modes that identify true dynamical features, and those which consist largely of noise. 

Simple ways of achieving this objective can involve either first projecting the data onto a smaller dimensional basis (such as the most energetic POD modes) before applying DMD, or by choosing only the most dynamically important DMD modes after applying DMD to the full data. 
One can also truncate the data to a dimension larger than the assumed dimension of the dynamics, and then apply a balanced truncation to the resulting dynamical system to obtain the desired reduced order model. This  approach is sometimes referred to as \emph{over-specification} in the system identification community \citep[see, e.g.,][]{juang1987eradc}. Keeping a higher dimension of data than that of the assumed dynamics can be particularly important for input-output systems that have highly energetic modes that are not strongly observable or controllable \citep{rowley:05pod}. Ideally, any algorithm that restricts the number of DMD modes that are computed should also additionally be computationally efficient. A fast method to perform DMD in real time on large datasets was recently proposed in \cite{Hemati2014streaming}, while
a library for efficient parallel implementation of number of common modal decomposition and system identification techniques is described in \cite{belson2013modred}.
An extension of DMD that potentially allows for better representation of nonlinear data has also recently been proposed \citep{williams2014edmd}, and although the computational costs increase dramatically with the dimension of the system, a kernel method described in \cite{williams2014kernel} reduces the cost to be comparable to standard DMD.

One of the major advantages of DMD over techniques such as global stability analysis, for example, is that it can be applied directly to data, without the need for the knowledge or construction of the system matrix, which is typically not available for experiments \citep{schmid2010DMD}. For this reason, analysis of the sensitivity of DMD to the type of noise typically found in experimental results is of particular importance.
The effects of noise on the accuracy of the DMD procedure was systematically investigated in the empirical study of \cite{duke2012error}, for the case of a synthetic waveform inspired by canonical periodic shear flow instabilities. 
More recently, \cite{pan2015accuracy} have extended this type of analysis to more complex data with multiple frequencies, as might be found in typical fluids systems.  
The present work builds upon these previous studies by analytically deriving an expression that explicitly shows how DMD should be affected by noise, for the case where the noise is assumed to be sensor noise that is uncorrelated with the dynamics of the system. 
Our analysis complements the ``noise-robust'' DMD formulation in \cite{hemati2015tls} by explicitly quantifying the influence of noise on DMD.
Further, while our analysis is consistent with the total least-squares formulation in \cite{hemati2015tls}, we use the insights gained from our analysis to develop alternative techniques to total least-squares DMD that may be preferable in certain applications. 
Ultimately, the availability of multiple ``noise-aware'' DMD algorithms allows the user to approach dynamical analysis of noisy data from multiple angles, thus garnering more confidence in the computations.
We note that the case of process noise, where noise can interact with the dynamics of the system, is also the subject of recent work \citep{bagheri2014noise}. 

Our analysis uses a recent characterization of DMD \citep{tu2014dynamic}, which highlights the connection of DMD to related techniques that are used in other communities for the extraction of dynamical information from data. Many linear system identification techniques are closely related in that they are based around singular value decomposition of a data matrix; aside from DMD there is the eigensystem realization algorithm \citep{ERA:1985} and balanced proper orthogonal decomposition \citep{rowley:05pod}, for example. Indeed, the origin of such an approach seems to date back to the work of \cite{Kalman:1965}.

In this work, we first show that the dominant effect of noise on DMD is often deterministic. This not only allows us to accurately predict its effect, but  also allows for a correction to be implemented to recover the noise-free dynamics. 
As well as directly correcting for the noise, we present two other modifications of DMD, that both are able to remove this bias without needing to know the noise characteristics. 
Sect.~\ref{sec:2} develops the theory that characterizes the effect of noise on DMD, which subsequently motivates the formulation of our modified algorithms, which we term \emph{noise-corrected} DMD (ncDMD), \emph{forward-backward} DMD (fbDMD) and \emph{total least-squares} DMD (tlsDMD). 
In Sect.~\ref{sec:resultssynth}, we analyze the performance of these algorithms on a number of synthetic data sets, which are corrupted by Gaussian white noise. 
We additionally investigate how the algorithms perform on data with both sensor and process noise. 
 In Sect.~\ref{sec:3}, we use numerical and experimental data from flow past a cylinder undergoing periodic vortex shedding, to demonstrate the utility of the proposed modifications of DMD for real fluids data. 
\section{Characterizing noise in dynamic mode decomposition} 
\label{sec:2}
This section details the methodology that is used to analyze the effect of noise in DMD. 
After introducing DMD in Sect.~\ref{sec:DMD}, the effect of sensor noise in the data  on the results of DMD is studied in Sect.~\ref{sec:DMDnoise}, which in particular shows that DMD is biased to sensor noise. 
Sects.~\ref{sec:ncDMD}--\ref{sec:tlsDMD} formulate three different modifications of the DMD algorithm that are designed to remove this bias. 

\subsection{Dynamic mode decomposition}
\label{sec:DMD}


DMD has undergone a number of formulations, interpretations and modifications since its inception. Common to all methods is the requisite collection and arrangement of data, summarized now. Suppose we collect snapshots of data $\bx_i$, which we assemble as columns in the data matrix $Z$. For fluids systems $\bx_i$ will typically be a velocity field snapshot, but more generally it is a vector of observations of an evolving dynamical system at a given time. From $Z$, we select all pairs of columns that are sampled at a time difference $\Delta t$ apart, and place them into the matrices $X$ and $Y$ (where the data in a given column of $Y$ was collected $\Delta t$ after the equivalent column of $X$). Note that if $Z$ consists of a sequential time-series of data, then $X$ and $Y$ are simply $Z$ with the last and first columns excluded, respectively. Let $X$ and $Y$ each be $n$ by $m$ matrices, so we have $m$ pairs of snapshots, each of size $n$. 
By not explicitly requiring a single time-series of data, we allow for larger or irregular time gaps between snapshot pairs, the concatenation of data from multiple time-series, and for the removal of any corrupted or spurious data. 
Recently, \cite{tu2014dynamic} proposed an interpretation of DMD modes and eigenvalues as the eigendecomposition of the matrix 
\begin{equation}
\label{eq:DMDdef}
A = Y X^{+},
\end{equation}
where $X^+$ denotes the Moore-Penrose pseudoinverse of a matrix $X$. While this is a succinct interpretation, and one which will be useful in the ensuing analysis, it is typically not an efficient (or even feasible) means of performing DMD (as discussed in \cite{tu2014dynamic}). This is especially true when $n \gg m$, 
which is often the case in high-dimensional fluids systems. Instead, since $X$ and $Y$ have rank at most $\min(m,n)$, it is typically more efficient to first project the data onto a subspace that is (at most) of this dimension. One way to do this is by projecting the original snapshots onto the POD modes of the data, which is implicitly done in all DMD algorithms. Note that the POD modes of $X$ are the columns of $U$ in the singular value decomposition $X = U\Sigma V^*$ (though typically POD is performed after first subtracting the temporal mean of the data, which is not done here). We present here a typical algorithm to compute DMD, that is most similar to that proposed in \cite{tu2014dynamic} as {\it exact} DMD. 
\begin{algorithm} [DMD]
\label{alg:1} ~
 \begin{enumerate}
\item Take the reduced singular value decomposition (SVD) of $X$, letting $X = U\Sigma V^*$. 
\item (Optional) Truncate the SVD by only considering the first $r$ columns of $U$ and $V$, and the first $r$ rows and columns of $\Sigma$ (with the singular values ordered by size), to obtain $U_r$, $\Sigma_r$, and $V_r$
\item Let $\tilde{A} \defeq U_r^*YV_r\Sigma_r^{-1}$
\item Find the eigenvalues $\mu_i$ and eigenvectors $w_i$ of $\tilde{A}$, with $\tilde{A}w_i = \mu_i w_i$,
\item Every nonzero $\mu_i$ is a DMD eigenvector, with a corresponding DMD mode given by $\varphi_i \defeq \mu_i^{-1} Y V_r \Sigma_r^{-1} w_i$.
\end{enumerate}
\end{algorithm}
This method is  similar to the original formulation in \cite{schmid2010DMD}, but for the fact that in step 5 the DMD modes are no longer restricted to lie within the column space of $X$. We also explicitly provide the optional step of truncating the SVD of $X$, which might be done if the system is known to exhibit low dimensional dynamics, or in an attempt to eliminate POD modes that contain only noise. 
We note that this is not the only means to reduce the dimension of the identified system dynamics, nor is it necessarily optimal. Indeed, \cite{wynn2013omd} develops a method that optimizes the projection basis in parallel while performing a DMD-like eigendecomposition. \cite{jovanovic2014dmdsp} takes a different approach, seeking a small number of nonzero modes from the full eigendecomposition that best approximate the system dynamics. An empirical comparison between these various dimensionality-reduction techniques will be given in Sect \ref{sec:MethodsCompare}. Note that the continuous eigenvalues $\lambda_{ci}$ of the system are related to the discrete time eigenvalues identified via DMD via $\lambda_{ci} = log(\mu_i)/\Delta t$. The growth rate $\gamma_i$ and frequency $\omega_i$ associated with DMD mode $\varphi_i$ are then given by $\lambda_{ci} = \gamma_i + i \omega_i$.

The matrix $\tilde{A}$ is related to $A$ in Eq.~\eqref{eq:DMDdef} by $\tilde{A}  = U_r^*AU_r$.
While $A$ can be viewed as an approximating linear propagation matrix in $\mathbb{R}^n$ (i.e., the space of original data vectors), $\tilde{A}$ is the equivalent propagation matrix in the space of POD coefficients, which we will sometimes refer to as POD space. Another interpretation of $\tilde{A}$ is that it is the spatial correlation matrix between the POD modes $U_r$, and the same POD modes shifted by the assumed dynamics $A$ \citep{schmid2010DMD}. If we let $\tilde{\bx}_k = U_r^*\bx_k$ be the representation of a given snapshot $\bx$ in the POD basis and let $\tilde{X}= U_r^*X$ and $\tilde{Y} = U_r^*Y$, then it is easy to verify that the equivalent of Eq.~\eqref{eq:DMDdef} in POD space is
 \begin{equation}
 \label{eq:DMDinPOD}
  \tilde{A} = \tilde{Y}\tilde{X}^+.
  \end{equation}
Eq.~\eqref{eq:DMDinPOD} will be useful for the subsequent analysis performed in this paper.

\subsection{Sensor noise in DMD}
\label{sec:DMDnoise}
In this work we use the term sensor noise to describe additive noise that affects only our measurements of a given system, and does not interact with the true dynamics. If we have a discrete-time dynamical system 
\begin{equation*}
\bx(t+\Delta t) = F[\bx(t)],
\end{equation*}
then we assume that our measurements take the form
\begin{equation*}
\bx_m(t) = \bx(t) + \bn(t),
\end{equation*}
where $\bn(t)$ is a random noise vector. For the purposes of this paper, we will take each component of $\bn(t)$ to be independent and normally distributed with zero mean and a given variance.
\todo{Should probably say a little more about what we assume about this random noise.
For instance, zero mean.  I don't think we need to be pedantic about the probability theory here, but probably good to be a little more specific than this.} 
With $X$ and $Y$ as described in Sect.~\ref{sec:DMD}, suppose that we measure $X_m = X + N_X$ and $Y_m = Y+ N_Y$, where $N_X$ and $N_Y$ are random matrices of sensor noise.
Note that some (or most) columns of random data in $N_X$ might also be in $N_Y$,
but shifted to a different column.
We will assume that the noise is independent of the true data, and is
independent in both space and time, so that each element of a given noise matrix is a random variable taken from a fixed zero-mean normal distribution.
%
From Eq.~\eqref{eq:DMDinPOD}, the measured DMD matrix $\tilde A_m$ can be computed from 
\begin{align}
\nonumber
\tilde A_m = \tilde Y_{m} \tilde X_{m}^+ &= ( \tilde Y + \tilde N_Y)( \tilde X + \tilde N_X)^+ \\
 \nonumber
 				&= ( \tilde Y + \tilde N_Y)( \tilde X + \tilde N_X)^* [( \tilde X + \tilde N_X)( \tilde X +\tilde  N_X)^*]^+ \\
				\label{eq:DMDnoisefull}
				&= (\tilde Y\tilde X^*+\tilde N_Y\tilde X^*+\tilde Y\tilde N_X^*+\tilde N_Y\tilde N_X^*) \left[ \tilde X\tilde X^*+\tilde N_X\tilde X^*+\tilde X\tilde N_X^*+\tilde N_X\tilde N_X^*\right]^+, 
\end{align}
where we have used the identity $M^+  = M^*(MM^*)^+$. Note that here the $\tilde{\cdot}$ notation means that the data is expressed in the POD basis obtained from the noisy data. 
We perform our analysis in this POD space rather than with the original data to allow for truncation of low energy modes, and because the computation of the pseudoinverse $X^+$ can be prohibitive for large datasets. 
We expect that the presence of noise should result in some error in the computation of $\tilde A_m$ (in comparison to the noise free matrix $\tilde A$) and thus some amount of error in the computed DMD eigenvalues and modes. Since elements of $\tilde A_m$ are statistical quantities dependent on the noise, it will make sense to compute statistical properties of the matrix. We begin by computing $\bE[\tilde A_m]$, the expected value of the computed DMD matrix. 
Provided that we have truncated any POD modes with zero energy, $\tilde X \tilde X^*$ should be invertible. If the noise terms are sufficiently small, then we can make use of the matrix perturbed inverse expansion $(M+P)^{-1} = M^{-1}-M^{-1}PM^{-1}+ \dots$, where higher order terms will be small for $M \gg P$. Eq.~\refeq{eq:DMDnoisefull} then becomes
{
\begin{equation}
\label{eq:DMDnoisefullexpanded}
\tilde A_m = (\tilde Y\tilde X^*+\tilde N_Y\tilde X^*+\tilde Y\tilde N_X^*+\tilde N_Y\tilde N_X^*)(\tilde X\tilde X^*)^{-1} 
\left[ I  - ( \tilde N_X\tilde X^*+\tilde X\tilde N_X^*+\tilde N_X\tilde N_X^*)(\tilde X\tilde X^*)^{-1}+\dots \right].
\end{equation} 
}

Taking the expected value of Eq.~\eqref{eq:DMDnoisefullexpanded}, we may classify the terms into three categories: a deterministic terms that does not involve $\tilde N_X$ or $\tilde N_Y$ (which ends up being $\tilde A$), terms involving one or three noise matrices, which will have expected values of $0$ (e.g., $\tilde N_Y \tilde X^* (\tilde X\tilde X^*)^{-1}$), and terms which involve two or four noise matrices. It is terms this latter category that may have non-zero expected values, and thus bias the result of applying DMD to noisy data. Discarding terms containing a single noise matrix, and additionally discarding higher order terms from the expansion, we have
\begin{align}
\bE(\tilde A_m)
\nonumber
 &= \tilde A(I - \bE( \tilde N_X\tilde N_X^* ) (\tilde X\tilde X^*)^{-1}) + \bE(\tilde N_Y\tilde X^+ \tilde N_X) \tilde X^+ 
 + \bE(\tilde N_Y\tilde X^+ \tilde X \tilde N_X^*)(\tilde X\tilde X^*)^{-1}\\
 \nonumber
 &  + \tilde Y\bE(\tilde N_X^* (\tilde X\tilde X^*)^{-1} \tilde N_X )\tilde X^+ + \tilde Y\bE(\tilde N_X^* (\tilde X\tilde X^*)^{-1} \tilde X\tilde N_X^* (\tilde X\tilde X^*)^{-1}) \\
 \label{eq:DMDbiasAllterms}
  & + \bE\left[\tilde N_Y \tilde N_X^* (\tilde X\tilde X^*)^{-1}(I  - \tilde N_X\tilde N_X^*(\tilde X\tilde X^*)^{-1})\right],
\end{align}
where we have noted that $\tilde Y \tilde X^*(\tilde X\tilde X^*)^{-1} = \tilde Y \tilde X^+ = \tilde A$. 
Assuming that the noise is sufficiently small compared with the true data, we can further neglect the term involving four noise matrices. The largest of the remaining terms will be that which contains the product $\tilde N_X \tilde N_X^*$. The remaining terms do not necessarily have zero mean, but for the purposes of this investigation will be neglected. 
Our results will demonstrate that this simplification is justifiable.  
This reduces Eq.~\eqref{eq:DMDbiasAllterms} to the following expression, relating the identified and true DMD matrices:
\begin{align}
\label{eq:DMDev}
\bE(\tilde A_m) &= \tilde A(I - \bE( \tilde N_X\tilde N_X^* ) (\tilde X\tilde X^*)^{-1} ). 
\end{align}
It might seem surprising that Eq.~\eqref{eq:DMDev} contains $N_X$, but not $N_Y$. 
 The reason for this will become apparent in Sect.~\ref{sec:tlsDMD}, where casting DMD in an optimization framework shows that the standard algorithm is optimal only when assuming that all of the noise is in $Y$, but not $X$.
 From a mathematical point of view, it is because the expression $\tilde A = \tilde Y \tilde X^+$ is linear in $\tilde Y$ but not in $\tilde X$, which is why perturbations to $\tilde X$ do not have to propagate through the equation in an unbiased manner.
 Note that the same analysis can be performed without transforming into POD space (i.e., without the $\tilde\cdot$ notation),  with the analogous expression to Eq.~\eqref{eq:DMDev} being
\begin{equation}
\bE( A_m) =  A(I - \bE(  N_X N_X^* ) ( X X^*)^{-1} ),
\end{equation}
subject to $X X^*$ being invertible. For systems where the size of each snapshot is larger than the number of snapshots (i.e, $n>m$, which is typical for fluids systems), $X X^*$ will not be invertible, thus motivating our choice to work in POD space. Moreover, one might want the option to truncate all but a certain number of POD modes, in order to obtain a low-dimensional model for the dominant system dynamics. Up until this point, we have not made a distinction between the POD modes of the clean data, $X$, and the noisy measured data, $X_m$, with the latter typically being all that we have access to. This issue will be explicitly addressed in Sect.~\ref{sec:ncDMD}. 
\begin{figure}
\center{\includegraphics[width= 0.35\textwidth]{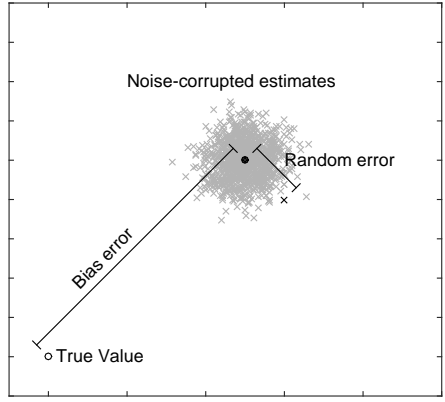}}
 \caption{Illustrative diagram showing how the error in estimation of a given quantity can be decomposed into bias error (being the difference between the true and expected value of the identified quantity), and random error (representing the fluctuation in the estimated quantity between different noise realizations)}
\label{fig:biasschematic}      
\end{figure}

Eq.~\eqref{eq:DMDev} shows that DMD is biased to sensor noise. 
In practice, the importance of this finding will depend on how the magnitude of this bias compares to the random component of error, that will fluctuate with different samples of random noise.
 Figure \ref{fig:biasschematic} shows an illustration of how bias and random components of error contribute to the total error in the estimation of some quantity from noisy data.
 Appendix 1 provides scaling arguments that suggest that the bias will be the dominant component of error in DMD whenever $m^{1/2}SNR > n^{1/2}$, where $SNR$ is the signal-to-noise ratio.
 When this is the case, it would be particularly advantageous if one had access to a bias-free alternative to DMD. The remainder of this section will present a number of such alternatives.

\subsection{Direct correction of sensor noise bias in DMD}
\label{sec:ncDMD}
Referring back to Eq.~\eqref{eq:DMDev},  we can form a bias-free estimate of the true DMD matrix $\tilde A$ via
\begin{equation}
\label{eq:nc1}
\tilde A \approx \tilde A_m( I -\bE(\tilde N_X \tilde N_X^*)(\tilde X \tilde X^*)^{-1})^{-1}.
\end{equation}
Making this modification in practice requires an accurate estimate of both the noise covariance, $\bE(\tilde N_X \tilde N_X^*)$, and the true data covariance,  $\tilde X \tilde X^*$, in POD space. 
For noise that is sufficiently small, we can utilize the approximation
\begin{equation} 
\label{eq:covapprox}
\tilde X \tilde X^* = U^* X X^* U \approx U^*_m X_m X_m^* U_m  =\Sigma_m^2,
\end{equation}
where $U_m\Sigma_mV_m^*$ is the singular value decomposition of the noisy data, $X_m$. This allows for us to express the bias of DMD in terms of  quantities that are measurable from noisy data. The assumption that $XX^*  = (X_m-N_X)(X_m-N_X)^*\approx X_m X_m^*$ can be further refined by retaining the $N_XN_X^*$ term, but for small noise this higher order term will typically be small enough to neglect after being inserted into Eq.~\eqref{eq:nc1}. 
The assumption that $U \approx U_m$ will largely be justified by means of results that show the utility of this analysis.
Analyzing the precise relationship between $U$ and $U_m$ in more detail is beyond the scope of this work, and is indeed an active area of research. We direct the interested reader to relevant results in perturbed SVD's  \citep{stewart1991perturbation,stewart2006perturbation,bai2009spectral}, random inner product matrices \citep{cheng2013spectrum, tao2012random}, and POD-type operations on noisy data \citep{epps2010error,singer2013two,zhao2013fourier}. 

Assuming that the noise is uniform as well as spatially and temporally independent, then $\bE( \tilde N_X\tilde N_X^* ) = \bE(U^*N_X N_X^*U) = U^*m \sigma_N^2U  = m \sigma_N^2 I$, where $\sigma_N^2$ is the variance of each independent component of the noise matrix.
With this assumption, and the approximation given in Eq.~\eqref{eq:covapprox}, Eq.~\eqref{eq:nc1}  becomes
 \begin{equation}
 \label{eq:noiseadjustmentPOD1}
\tilde A \approx \tilde A_m( I -m\sigma_N^2\Sigma_m^{-2})^{-1}.
\end{equation}
If the noise is sufficiently small, then a perturbed inverse approximation gives
\begin{equation}
\label{eq:noiseadjustmentPOD2}
\tilde A \approx \tilde A_m(I +m\sigma_N^2\Sigma_m^{-2}).
\end{equation}
We thus have derived a correction to the bias that is present in the original DMD matrix $A_m$ due to the effect of sensor noise. We note that this approximation relies on an accurate knowledge of the noise covariance matrix. There are numerous means to estimate noise properties from data, see \cite{Pyatykh2013image}, for example. The approximations used in deriving this expression also rely on the magnitude of the noise being smaller than that of the true data within each non-truncated POD mode.
We now state explicitly the algorithm by which we can correct for the effect of sensor noise in the DMD algorithm, which we refer to as noise-corrected DMD, or ncDMD:
\begin{algorithm}[Noise-corrected DMD (ncDMD)]
\label{alg:2}~
\begin{enumerate}
\item Compute $\tilde{A}_m$ from the measured data as per steps $1$--$3$ of Algorithm \ref{alg:1}
\item Compute the approximation of $\tilde{A}$ from Eq.~\eqref{eq:noiseadjustmentPOD1} 
\item Compute the DMD eigenvalues and modes via steps $4$--$5$ of Algorithm \ref{alg:1}, using the bias-free estimate of $\tilde{A}$.
\end{enumerate}
\end{algorithm}
As was also noted in Sect \ref{sec:DMDnoise}, we could have performed all of the above analysis without first projecting onto the space of POD coefficients, which gives us the following as analogous to Eqs.~\eqref{eq:noiseadjustmentPOD1} and~\eqref{eq:noiseadjustmentPOD2} respectively, subject to the appropriate inverses existing:
\begin{equation}
A \approx A_m(I - m\sigma_N^2(X_mX_m^*)^{-1} )^{-1} \approx  A_m(I + m\sigma_N^2(X_mX_m^*)^{-1} ).
\end{equation}
While this approach might be computationally prohibitive for many applications of DMD (since it requires inverting large $n\times n$ matrices), it could in theory be more accurate, since it doesn't rely on any assumption that the POD modes for the measured and true data are sufficiently close to each other. Note again that $XX^*$ can only be invertible if $m > n$, as otherwise it cannot be full rank.

\subsection{Forward-backward DMD}
\label{sec:fbDMD}
If we were to swap the data in $X$ and $Y$, then (for suitably well behaved data) we should expect to identify the inverse dynamical system, with state propagation matrix $B_m$ (or $\tilde B_m$ in POD space), which approximates the true dynamics $B$ (and $\tilde B$). Note that it is not guaranteed that the dynamics of the original system are invertible, but this assumption should not be too restrictive for the majority of physical systems under consideration (particularly after projection onto an appropriate POD subspace).
It is argued in Appendix 1 that sensor noise has the effect of shifting the computed DMD eigenvalues to appear to be more stable than they actually are (i.e., moving them further inside the unit circle). Since our analysis was independent of the nature of the data, we should expect the same effect to be present for the computation of the inverse system. However, if $\tilde B$ is invertible, then we should have $\tilde B = \tilde A^{-1}$, meaning that we should be able to compute an estimate of the forward-time propagation matrix using backward-time DMD, via $\tilde A_m^{back} =  \tilde B_m^{-1}$. However, given that the eigenvalues of $\tilde B_m$ should have their growth rates underestimated, those of the eigenvalues of $\tilde A_m^{back}$ will be overestimated. Specifically, from consideration of Eq.~\eqref{eq:DMDev}, we have
\begin{equation*}
\bE(\tilde B_m) = \tilde B\left(I - \bE( \tilde N_X\tilde N_X^* ) (\tilde X\tilde X^*)^{-1} \right),
\end{equation*}
and so
\begin{equation}
\tilde A_m^{back}\approx \left(I - \bE( \tilde N_X\tilde N_X^* ) (\tilde X\tilde X^*)^{-1} \right)^{-1}\tilde A,
\end{equation}
where we are using the fact that the noise and POD energy components are the same for forward- and backward-DMD. 
We can then combine estimates of the dynamics from forward- and backward-time DMD to obtain
\begin{equation}
\tilde A_m \tilde A_m^{back} = \tilde A \left(I - \bE( \tilde N_X\tilde N_X^* ) (\tilde X\tilde X^*)^{-1} \right) \left(I - \bE( \tilde N_X\tilde N_X^* ) (\tilde X\tilde X^*)^{-1} \right)^{-1}\tilde A = \tilde A^2.
\end{equation}
We thus have the estimate
\begin{equation}
\label{eq:geomAv}
\tilde A \approx (\tilde A_m \tilde A_m^{back})^{1/2}.
\end{equation}
Note that this square root will in general be non-unique, and thus determining which root is the relevant solution could be nontrivial. One reasonable method, if there is any ambiguity, is to take the square root which is closest to $\tilde A_m$ (or $\tilde A_m^{back}$). See \cite{golub2012matrix} for a more detailed discussion of the computation of matrix square roots. As an aside, note that if we assume that we know the equivalent continuous time matrices $\tilde A^c_m = \log (\tilde A_m)/\Delta t$ and   $\tilde A^{c,back}_m = \log (\tilde A^{back}_m)/ \Delta t$, then the equivalent of Eq.~\eqref{eq:geomAv} is
\begin{equation*}
\tilde A^c \approx \frac{1}{2}(\tilde A^c_m + \tilde A_m^{c,back}).
\end{equation*}
We are now in a position to formalize this algorithm, which we refer to as \emph{forward-backward} DMD or fbDMD.
\begin{algorithm}[forward-backward DMD (fbDMD)]
\label{alg:3} ~
\begin{enumerate}
\item Compute $\tilde{A}_m$ from the measured data as per steps $1$--$3$ of Algorithm \ref{alg:1}
\item Compute $\tilde{B}_m$ from the measured data as per steps $1$--$3$ of Algorithm \ref{alg:1}, where $X$ and $Y$ are interchanged
\item Compute the approximation of $\tilde{A}$ from Eq.~\eqref{eq:geomAv} 
\item Compute the DMD eigenvalues and modes via steps $4$--$5$ of Algorithm \ref{alg:1}, using the improved estimate of $\tilde{A}$ from step 4.
\end{enumerate}
\end{algorithm}
Note that in the case where most data snapshots are in both $X$ and $Y$ (e.g., for a sequential time series of data) we can reduce the computational cost of steps 1--2 in Algorithm \ref{alg:3} by first taking the SVD of the entire data set, and then working is the space of the resulting POD modes.

\subsection{Total least-squares DMD}
\label{sec:tlsDMD}
For the case where the number of snapshots, $m$, is greater than the size of each snapshot, $n$, the DMD matrix $A$ can be interpreted as the least-squares solution to the overdetermined system $A X = Y$. When $n > m $, then the solution for the now underdetermined system is the minimum Frobenius norm solution to $A X = Y$. In both cases, this solution is $A  = Y X^+$. Note that it is possible to turn an under-determined system into an over-determined system by truncating the number of POD modes used to less than $m$ (truncating to precisely $m$ results in a unique solution when the data is full column rank, with no loss of data). A least-squares solution of this form minimizes the error in $Y$, but implicitly assumes that there is no error in $X$. 
 This can explain why the bias in DMD (Eq.~\eqref{eq:DMDev}) is dependent on $\tilde N_X$, but not $\tilde N_Y$.
That is, in the least-squares case DMD can be viewed as finding
\begin{equation*}
A: \  Y + E_Y = AX, \ \ \text{minimizing} \ \|E_Y\|_F ,
\end{equation*}
where $\|\cdot\|_F$ denotes the Frobenius norm of a matrix. When doing backwards-time DMD in Sect.~\ref{sec:fbDMD}, we conversely assume that $Y$ is known exactly and minimize the error in $X$. That is, assuming the identified dynamics are invertible, we find
\begin{equation*}
A: \  Y  = A(X+E_X) , \ \ \text{minimizing} \ \|E_X\|_F  .
\end{equation*}
 For this reason, combining forward- and backward-time DMD takes into account the error in both $X$ and $Y$.  A more direct means of doing this is to use a single algorithm that finds a least-squares solution for the error in both $X$ and $Y$. It is possible to adapt standard TLS algorithms \citep{golub2012matrix} to a DMD setting, which we perform here.
 We seek
\begin{equation*}
A:  \  (Y+E_Y)  = A(X+E_X), \ \ \text{minimizing} \  \|E\|_F, \  \text{where} \  E = \begin{bmatrix} E_X \\ E_Y \end{bmatrix} .
\end{equation*}
The expressions $Y+E_Y$ and $X+E_X$ can be interpreted as $Y_m-N_Y$ and $X_m-N_X$. 
To solve for this, we can rearrange the equation to obtain 
\begin{equation}
\label{eq:TLS1}
\begin{bmatrix} A &  -I \end{bmatrix}   \begin{bmatrix} X+ E_X \\ Y+ E_Y \end{bmatrix}  = 0.
\end{equation}
We would now like to assume that $2n < m$. This might not be the case, particularly for high-dimensional fluids data. To get around this, and improve computational tractability, we may project Eq.~\eqref{eq:TLS1} onto a POD subspace of dimension $r < m/2$, to obtain
\begin{equation}
\label{eq:TLS1POD}
\begin{bmatrix} \tilde A &  -I \end{bmatrix}   \begin{bmatrix} \tilde X+\tilde  E_X \\ \tilde Y+ \tilde E_Y \end{bmatrix}  = 0.
\end{equation}
This POD projection step is in contrast to the TLS DMD formulation in \cite{hemati2015tls}, where a projection is performed onto a basis determined from an augmented snapshot matrix $Z=\begin{bmatrix}X \\ Y  \end{bmatrix}$.
We find that the present formulation yields more accurate eigenvalues in a number of examples. Note that the nullspace of $\begin{bmatrix} \tilde A &  -I \end{bmatrix}$ is $r$-dimensional, meaning that the $2r$ by $m$ matrix $ \begin{bmatrix} \tilde X+ \tilde E_X \\ \tilde Y+\tilde E_Y \end{bmatrix}$ can have rank at most $r$.  

 Let the full SVD of  $\begin{bmatrix}\tilde  X \\ \tilde Y  \end{bmatrix}$ be given by $U\Sigma V^*$. If the data is noisy, we should expect that all $2r$ diagonal entries of $\Sigma$ are nonzero. By the Eckart-Young theorem \citep{eckart1936approx}, the nearest (in the sense of Frobenius norm) rank $r$ matrix will be given by
\begin{equation*}
\begin{bmatrix} \tilde X+\tilde  E_X \\ \tilde Y+ \tilde E_Y \end{bmatrix} = U \Sigma_{1:r} V^*,
\end{equation*}
where $\Sigma_{1:r}$ contains the leading $r$ singular values of $\Sigma$, with the rest replaced by zeros. 
We then have that
\begin{equation*}
\begin{bmatrix} \tilde X+\tilde  E_X  \\  \tilde Y+\tilde  E_Y \end{bmatrix} =  U \Sigma_{1:r}V^* 
 = \begin{bmatrix} U_{11} & U_{12} \\ U_{21} & U_{22}\end{bmatrix} 
 \begin{bmatrix}  \Sigma_1 & 0 \\ 0 & 0 \end{bmatrix}\begin{bmatrix}V_1^* \\ V_2^*\end{bmatrix} 
 = \begin{bmatrix}  U_{11}\Sigma_1 V_1^* \\ U_{21}\Sigma_1 V_1^*  \end{bmatrix},
 \end{equation*}
where $U_{ij}$ are $r$ by $r$ matrices, and $V_1$ is the first $r$ columns of $V$. Rearranging this equation, we obtain the total least-squares estimate for $\tilde A$:
\begin{equation}
\label{eq:TLS3}
\tilde A = U_{21} U_{11}^{-1}.
\end{equation}
Note that this derivation requires that $U_{11}$ be invertible. While the derivation includes the full SVD of the augmented data, Eq.~\eqref{eq:TLS3} indicates that we only need the first $r$ columns of $U$, meaning that only a reduced SVD is required.  Algorithm~\ref{alg:4} summarizes this total least-squares approach to DMD, which we refer to as \emph{total least-squares} DMD, or tlsDMD. 
\begin{algorithm}[total least-squares DMD (tlsDMD)]
\label{alg:4} ~
\begin{enumerate}
\item Collect data $X$ and $Y$, and project onto $r < m/2$ POD modes to obtain $\tilde X$ and $\tilde Y$.
\item Take the SVD of $\begin{bmatrix} \tilde X \\ \tilde Y\end{bmatrix}$, letting $\begin{bmatrix}\tilde  X \\ \tilde Y\end{bmatrix} = U\Sigma V^*$. 
\item Partition the $2r$ by $2r$ matrix $U$ into $r$ by $r$ sub-matrices, letting $U = \begin{bmatrix}U_{11} & U_{12} \\ U_{21} & U_{22}\end{bmatrix}$ (note that only the first $r$ columns need to be computed).
\item Compute the total least-squares DMD matrix $\tilde A$, using Eq.~\eqref{eq:TLS3}.
\item Compute the DMD eigenvalues and modes using steps $4$--$5$ of Algorithm \ref{alg:1}.
\end{enumerate}
\end{algorithm}
An alternative and more focused exposition of tlsDMD is given in \cite{hemati2015tls}.  We note that Algorithm~\ref{alg:4} is not identical to that presented in this work (due to the lack of pre-truncation of POD modes), however we find that Algorithm~\ref{alg:4} gives marginally better results in terms of the accuracy of identified eigenvalues.

\section{Results with synthetic data}
\label{sec:resultssynth}
In this section we will test our proposed modifications to DMD on a number of examples.  Using known dynamics with the addition of random noise will allow us to examine the performance of these proposed modifications (Algorithms \ref{alg:2}--\ref{alg:4}) in comparison to regular DMD (Algorithm \ref{alg:1}). 
We begin by considering a simple 2-dimensional linear system in Sect.~\ref{sec:ex1}. In Sect.~\ref{sec:ex2}, we consider the same system with an expanded set of observables, which tests the important case of high-dimensional data that is described by low-dimensional dynamics. Sect.~\ref{sec:MethodsCompare} compares the performance of the proposed modifications of DMD to other DMD variants in recent literature, while Sect.~\ref{sec:hidden} considers the problem of identifying dynamics that are quickly decaying and obscured by dominant modes and noise, a case where DMD-like algorithms could be of most use.
Finally, in Sect.~\ref{sec:process} we analyze how the proposed DMD modifications treat process noise.

\subsection{Example: A periodic linear system} \label{sec:ex1}
We consider first a simple two-dimensional linear system, with dynamics given by 
\begin{equation}
\label{eq:ex1}
\dot\bx = \begin{bmatrix}1 & -2 \\
					1& -1 \end{bmatrix} \bx.
\end{equation}

\begin{figure*}
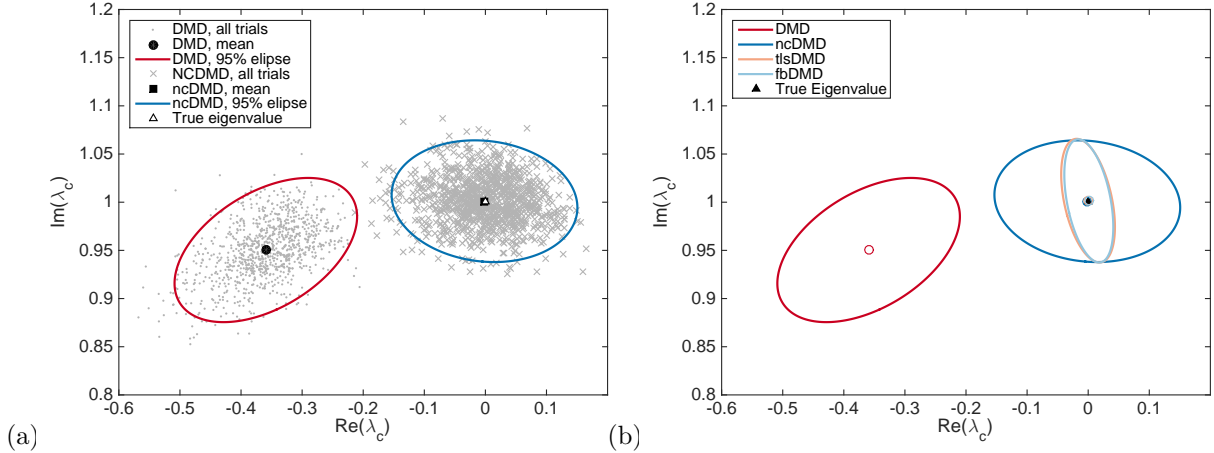

(a)  \includegraphics[width= 0.45\textwidth]{Fig2a.eps}(b) \includegraphics[width= 0.45\textwidth]{Fig2b.eps}
\caption{(a) Eigenvalues (in continuous-time) identified from regular DMD (Algorithm \ref{alg:1}, dots) and noise-corrected DMD (Algorithm \ref{alg:2}, crosses) from 100 snapshots of data from Eq.~\eqref{eq:ex1}, with $\Delta t = 0.1$ and $\sigma_N^2 = 0.01$. Only one of the complex conjugate pair of eigenvalues is shown. The mean and $95\%$ confidence ellipse of 1000 trials are given for each data set. (b) shows the mean and $95\%$ confidence ellipse for the same data set for Algorithms \ref{alg:1}--\ref{alg:4}}
\label{fig:eigs1}     
\end{figure*}
This system has (continuous-time) eigenvalues $\lambda_{c 1,2} = \pm i$, so gives purely periodic dynamics, with no growth or decay. We discretize with a timestep $\Delta t = 0.1,$ so the discrete-time eigenvalues are then $\lambda_{1,2} = e^{\pm\Delta t i}$. We use 100 timesteps of data (i.e., $m = 99$), corrupted with Gaussian white noise of variance $\sigma_N^2 = 0.01$. The identified continuous-time eigenvalues from both regular DMD (Algorithm \ref{alg:1}), and the direct noise-correction (Algorithm \ref{alg:2}) are shown in Fig.~\ref{fig:eigs1}(a), for $1000$ different trials from the initial condition $\bx_0 = [1 \  0.1]^T$. We assume that the correction term is given by $m\sigma_N^2 I_n$, and observe that this corrects almost perfectly for the bias in the DMD algorithm in terms of identifying eigenvalues. Also shown in Fig.~\ref{fig:eigs1}(a) are ellipses representing the $95\%$ confidence region, with the major and minor axes of the ellipse aligned with the principal component directions of the eigenvalue data. For clarity, in the presentation of subsequent results, we will omit individual data points and show only such ellipses. In Fig.~\ref{fig:eigs1}(b) we show the mean and $95\%$ confidence ellipses for Algorithms \ref{alg:3} and \ref{alg:4}. As with ncDMD, both fbDMD (Algorithm \ref{alg:3}) and tlsDMD (Algorithm \ref{alg:4}) accurately correct for the bias in the mean of the identified eigenvalue. Further to this, fbDMD and tlsDMD also both reduce the area of the 95\% confidence ellipse, which indicates that they are more likely to attain a closer approximation to the correct eigenvalue on any given trial. 

\begin{figure}
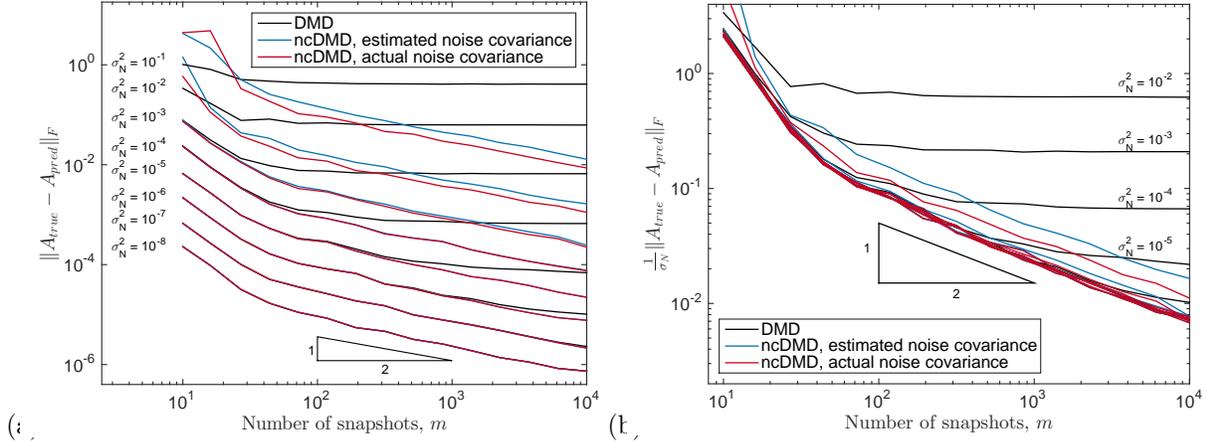

  (a)\includegraphics[width= 0.45\textwidth]{Fig3a.eps}
  (b)\includegraphics[width= 0.45\textwidth]{Fig3b.eps}
\caption{Error in the estimated propagation matrix $\tilde A$ arising from performing DMD and ncDMD on noise-corrupted data generated from Eq.~\eqref{eq:ex1}, for different values of $m$ and $\sigma_N^2$. In (a) the error is given as $\| A_{\text{true}}-A_{\text{pred}}\|_F$, while in (b) this quantity is normalized by the standard deviation of the noise, $\sigma_N$. In both cases, the error is averaged over 100 trials for each $m$ and $\sigma_N^2$. Note that for clarity, (b) excludes the two largest noise levels shown in (a)}
\label{fig:froberrors}      
\end{figure}

Focusing back on comparing Algorithms \ref{alg:1} and \ref{alg:2}, we show results for a variety of values of $m$ and $\sigma_N^2$ in Fig.~\ref{fig:froberrors}. In Fig.~\ref{fig:froberrors}(a), rather than looking at the error in the eigenvalues, we instead consider the Frobenius norm of the difference between the true and identified propagation matrices, $\| A_{\text{true}}-A_{\text{pred}}\|_F$. 
For very small noise, the correction makes little difference, since the random error is larger than the bias error. 
For larger values of noise, we observe that the error saturates when using standard DMD, which is due to the presence of the bias term identified in Sect.~\ref{sec:DMDnoise}, which has a size independent of the number of samples, $m$. We note that the magnitude of this bias term is proportional to $\sigma_N^2$, as predicted by Eq.~\eqref{eq:noiseadjustmentPOD1}. 
Evidence of this error saturation phenomena can also be seen in past studies of the effect of sensor noise on DMD \citep{duke2012error,wynn2013omd,pan2015accuracy}.
 After this bias term is corrected for, we see that the error decays proportional to $m^{-1/2}$ for all values of noise, as predicted from the analysis in Appendix 1. 
 The more rapid decay in error with $m$ for small numbers of samples seems to arise from the fact that the data has not yet completed one full period of oscillation. Fig.~\ref{fig:froberrors} shows the corrections to DMD made using both the sampled ($N_XN_X^*$) and theoretical ($m\sigma_N^2 I$) covariance matrices. 
 Normally the sample noise covariance would not be known, and so we demonstrate here that the theoretical covariance achieves almost the same decrease in error. 
 Fig.~\ref{fig:froberrors}(b) shows that the ncDMD error curves collapse when the error is normalized by the standard deviation of noise, $\sigma_N$ (note that we could also multiply the error by the $SNR$ to get the same scaling). 

\begin{figure}
\center{\includegraphics[width= 0.45\textwidth]{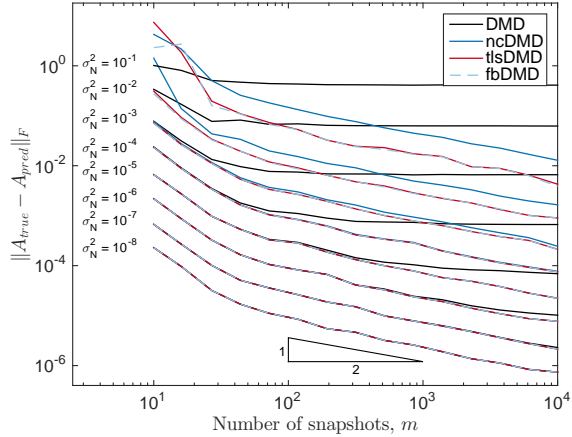}}
\caption{Error in the estimated propagation matrix $\tilde A$ arising from performing DMD, ncDMD, fbDMD, and tlsDMD on noise-corrupted data generated from Eq.~\eqref{eq:ex1}, for different values of $m$ and $\sigma_N^2$. The error is given as $\| A_{\text{true}}-A_{\text{pred}}\|_F$, and is averaged over 100 trials for each $m$ and $\sigma_N^2$}
\label{fig:froberrorsmore}      
\end{figure}

Fig.~\ref{fig:froberrorsmore} shows the performance of Algorithms \ref{alg:3} and \ref{alg:4} on the same data as Fig.~\ref{fig:froberrors}. Again, we find that both of these algorithms can prevent the error saturation present in standard DMD, and indeed can perform noticeably better than Algorithm \ref{alg:2} for larger noise levels. Algorithms \ref{alg:2}--\ref{alg:4} all appear to exhibit the same asymptotic behavior as the number of snapshots, $m$, increases, with the error decreasing proportional to $m^{-1/2}$.

A common means to mitigate the effect of noisy data is to collect multiple time-series of data, and process this in such as way to improve the results over just using one data set. One can ask the question if it is better to concatenate the snapshots of data from each time series and apply DMD to this collection, or to apply DMD to phase-averaged data. Our results suggest that the latter option is preferable if using standard DMD, since adding additional pairs of snapshots will not decrease the error beyond a certain level, due to this bias saturation at large $m$. If we are using ncDMD, fbDMD, or tlsDMD, however, then we should get the same scalings regardless of which option is chosen, since in both cases the error should be proportional to $p^{-1/2}$, where $p$ is the number of trials of data collected.

\subsection{A periodic linear system with a high-dimensional state of observables} \label{sec:ex2}

This example considered in Sect.~\ref{sec:ex1} has $m\gg n$, which is atypical of many fluids systems for which DMD is used. To consider the case where the size of the state $n$ is larger than the number of snapshots $m$, we expand the state of our system to include time-shifts of the data. In this sense, we have new observables given by
\begin{equation}
\label{eq:ex1X}
\bz_k = \left[\begin{array}{c}\bx_k \\\bx_{k-1} \\\vdots \\\bx_{k-q}\end{array}\right],
\end{equation}
with the size of the state $n = 2(q+1)$. This periodic system can equivalently be viewed as a traveling wave, which is now observed over a larger spatial domain. Similar data (but with a non-zero growth rate) was considered in \cite{duke2012error} and \cite{wynn2013omd}. Since the dynamics are still only two-dimensional despite the higher dimensional state, we use only the first two POD modes of the data to identify a $2\times2$ propagation matrix $\tilde A$. The next section will examine alternative means of performing this dimensionality reduction. 

\begin{figure}
\center{\includegraphics[width= 0.48\textwidth]{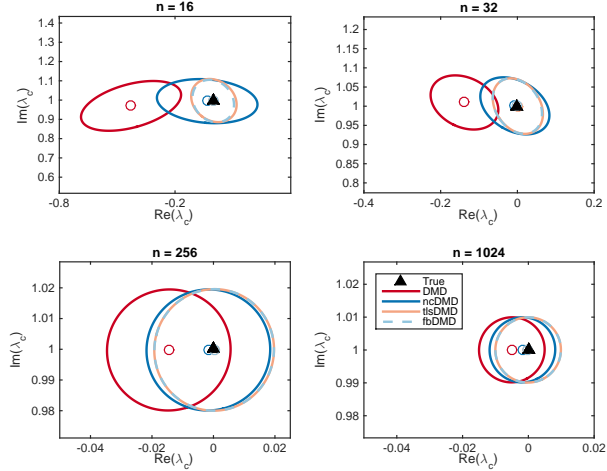}}
 \caption{Mean and $95\%$ confidence ellipses of continuous-time eigenvalues identified by applying regular DMD (Algorithm \ref{alg:1}), noise-corrected DMD (Algorithm \ref{alg:2}), forward-backward DMD (Algorithm \ref{alg:3}) and total least-squares DMD  (Algorithm \ref{alg:4}) for noisy data generated from 1000 trials of data generated by Eq.~\eqref{eq:ex1}, and observed as in Eq.~\eqref{eq:ex1X}. Here the number of snapshots $m$ is fixed to be 50, $\Delta t = 0.1$, and $\sigma_N^2 = 0.1$. %
  Only one of the complex conjugate pair of eigenvalues is shown}
\label{fig:eigsX}      
\end{figure}

Fig.~\ref{fig:eigsX} shows the statistical results (in terms of DMD eigenvalues) of performing variants of DMD on such data, using $m = 50$ and a range of snapshot sizes, $n$. We find that a bias exists for regular DMD, but the magnitude of this bias decreases as the size of each snapshot increases (note that the scale between subplots, thought the aspect ratio remains the same). We find that Algorithms \ref{alg:2}--\ref{alg:4} all outperform regular DMD in terms of giving mean (expected) eigenvalues that are closer to the true value. For small state sizes, Algorithms \ref{alg:3} and \ref{alg:4} also also give a smaller confidence ellipse, though this is not observed for larger state sizes.  As the size of the state increases, the bias component of the error of DMD (evidenced by the difference between the true and mean identified eigenvalue) becomes smaller relative to the random component of the error (indicated by the size of the confidence ellipse). This means that the modifications to DMD presented in Algorithms \ref{alg:2}--\ref{alg:4} give the largest improvement when the size of the state is small, due to the fact that in this regime the bias component of error is larger than the random component. Note that these conclusions may be predicted from the scaling laws given in Eqs.~\eqref{eq:biasscale} and~\eqref{eq:randomerror}. Moreover, one can verify that as the size of the state ($n$) increases, the size of the ellipses decrease proportional to $n^{-1/2}$.

\subsection{Comparison to other modified DMD algorithms}
\label{sec:MethodsCompare}
Without any modification, applying DMD on noisy data will give $\min(m,n)$ eigenvalue-mode pairs, many of which may be mostly or entirely due to noise, particularly if the underlying dynamics are low-dimensional. For this reason, a number of modifications of DMD that aim to identify a small number of dynamically important modes have been developed. The most simple means of reducing the dimension of the data is to simply project onto a reduced number of POD modes, which is explicitly mentioned as an optional step in Algorithm \ref{alg:1}. This projection step was also used within Algorithms \ref{alg:2}--\ref{alg:4} in Sect.~\ref{sec:ex2}. A number of alternative means to obtain a small number of dynamic modes from DMD-type algorithms have been proposed, as briefly mentioned in Sect.~\ref{sec:intro}. These variants all start with the observation that standard DMD can be formulated within an optimization framework, in the sense that it identifies a least-squares or minimum-norm propagation matrix for a given data set.  \cite{chen2011variants} proposes a modification termed \emph{optimized} DMD that seeks to find optimal low-rank dynamics that best matches a sequential time-series of data. While the fact that this method optimizes over the entire time-sequence of data rather than just pairwise snapshots should increase its robustness to noise, the non-convexity of the optimization potentially limits its utility.   Optimal Mode Decompostion (OMD, \cite{goulart2012omd,wynn2013omd}) finds an optimal low-dimensional subspace on which the identified dynamics reside, rather than assuming that this subspace is simply the most energetic POD modes. This approach was shown to give an improvement on the DMD eigenvalues obtained for noisy data in \cite{wynn2013omd}. Sparsity-promoting DMD (spDMD, \cite{jovanovic2014dmdsp}) adds an $l_1$ regularization term that penalizes the number of DMD modes with non-zero coefficients in the approximation of the time-series of data.  

\begin{figure}
\center{\includegraphics[width= 0.48\textwidth]{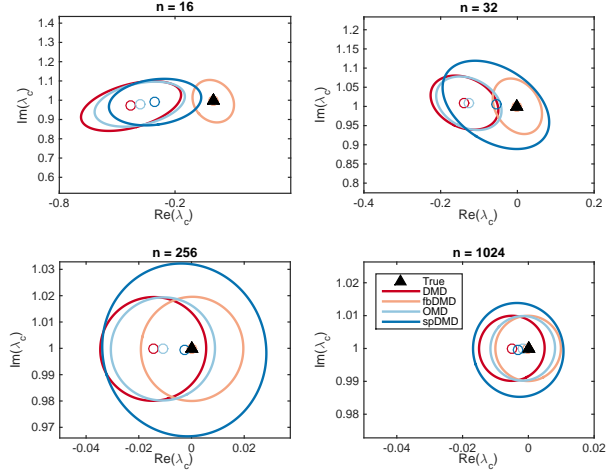}}
 \caption{Mean and $95\%$ confidence ellipses of continuous-time eigenvalues identified by applying regular DMD (Algorithm \ref{alg:1}), forward-backward DMD (Algorithm \ref{alg:4}), OMD and spDMD for noisy data generated from 1000 trials of data generated by Eq.~\eqref{eq:ex1}, and observed as in Eq.~\eqref{eq:ex1X}. Here the number of snapshots $m$ is fixed to be 50, $\Delta t = 0.1$, and $\sigma_N^2 = 0.1$. Only one of the complex conjugate pair of eigenvalues is shown}
\label{fig:MethodsX}      
\end{figure}

This section will compare OMD and spDMD with the algorithms presented in the present work. Of the algorithms presented here, we will  focus on fbDMD (Algorithm \ref{alg:3}), which was found to perform equally well as tlsDMD, and better than ncDMD, in Sects.~\ref{sec:ex1} and \ref{sec:ex2}. Fig.~\ref{fig:MethodsX} shows identified eigenvalue statistics (mean and confidence ellipses) for each of these algorithms, using the same data as that for Fig.~\ref{fig:eigsX}. We observe that OMD gives a more accurate mean eigenvalue that DMD, and a confidence ellipse of approximately the same size. spDMD gives a mean identified eigenvalue that is closer again to the mean, although the variance in the eigenvalues identified for each trial is larger.  We note that spDMD occasionally produced erroneous results, which were excluded as outliers from the statistical analysis. This highlights an important advantage to the modifications to DMD presented here - the algorithms are given in closed form, and do not rely on an appropriate selection of parameters and tolerances that are most likely required for an optimization procedure. In all of the cases, fbDMD (and tlsDMD, which is not shown but barely distinguishable from fbDMD) gives the best estimate of the true eigenvalue.

While these results suggest that fbDMD/tlsDMD is more accurate than OMD and spDMD, we must remember that the results from one data set do not show the global superiority of any given algorithm. Indeed, one could most likely find data sets for which any given algorithm is superior (by some chosen metric) to others.
We conclude this section by noting that it should be possible to combine the optimization procedures presented in  \cite{chen2011variants}, \cite{wynn2013omd}, and \cite{jovanovic2014dmdsp})
with the modifications to DMD presented here. Indeed, a simple means to do this might be to modify Algorithm \ref{alg:3} so that the results of applying a given algorithm forwards and backwards in time are geometrically averaged, as in Eq.~\eqref{eq:geomAv}.

\subsection{Identifying hidden dynamics}
\label{sec:hidden}
\begin{figure}
\center{\includegraphics[width= 0.48\textwidth]{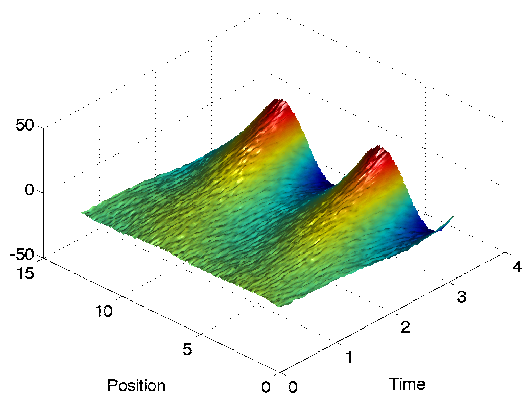}}
 \caption{Visualization of data generated by Eq.~\eqref{eq:hiddensig}, with $k_1 = 1, \omega_1 = 1,$ $\gamma_1$  =1, $k_2 = 0.4, \omega_2= 3.7$, $\gamma_2 = -0.2$, and $\sigma = 0.5$}
\label{fig:hiddenSurf}      
\end{figure}

\begin{figure}
\center{\includegraphics[width= 0.8\textwidth]{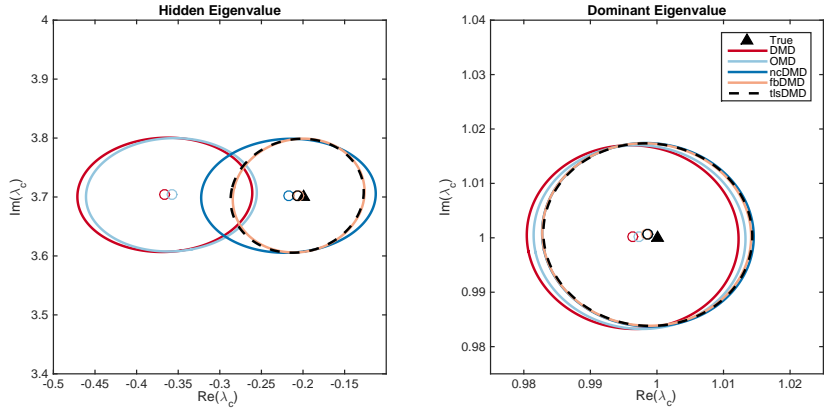}}
 \caption{Mean and $95\%$ confidence ellipses of continuous-time eigenvalues identified by applying regular DMD (Algorithm \ref{alg:1}), OMD, noise-corrected DMD (Algorithm \ref{alg:2}), forward-backward DMD (Algorithm~\ref{alg:3}) and total least-squares DMD  (Algorithm~\ref{alg:4}) to 1000 trials of noisy data generated by Eq.~\eqref{eq:hiddensig}, with $k_1 = 1, \omega_1 = 1,$ $\gamma_1$  =1, $k_2 = 0.4, \omega_2= 3.7$, $\gamma_2 = -0.2$, and $\sigma = 0.5$}
\label{fig:hiddenEigs}      
\end{figure}

The systems considered in Sects.~\ref{sec:ex1} and \ref{sec:ex2} could be considered ``easy'' in the sense that the dominant dynamics are simple, and of consistently larger magnitude than the noise. Indeed, it is not difficult to qualitatively identify such dynamics by eye from simply looking at some visualization of the data. 
A more difficult case occurs when some of the dynamics are of low magnitude and/or are quickly decaying, and thus might quickly be lost among the noise in the measurements. A major benefit of data processing techniques such as DMD is the ability to identify dynamics that might otherwise remain hidden. With this in mind, we now consider a superposition of two sinusoidal signals that are traveling across a spatial domain in time, with the amplitude of one mode growing, and the other decaying:
\begin{equation}
f(x,t) = \sin(k_1 x - \omega_1 t)e^{\gamma_1 t}+ \sin(k_2 x - \omega_2 t)e^{\gamma_2 t} + n_{\sigma}(x,t),
\label{eq:hiddensig}
\end{equation}
where we set $k_1 = 1,$ $\omega_1 = 1,$ $\gamma_1  =1$, $k_2 = 0.4$, $\omega_2= 3.7$ and $\gamma_2 = -0.2$. We thus have the superposition of a growing, traveling wave, and a decaying signal that is quickly hidden by the unstable dynamics. The four continuous-time eigenvalues of this data are $\gamma_1 \pm \omega_1$ and $\gamma_2 \pm \omega_2$. This data is again similar to that considered in \cite{wynn2013omd} and \cite{duke2012error}, if we neglect the decaying dynamics. 
Fig.~\ref{fig:hiddenSurf} shows the data with white noise of standard deviation $\sigma = 0.5$. 
Fig.~\ref{fig:hiddenEigs} shows the performance of various DMD-type algorithms in identifying one of the dominant eigenvalues ($1+i$) and one of the ``hidden'' eigenvalues ($-0.2 + 3.7i$). Mean eigenvalues and error ellipses are computed from 1000 different noise samples. Unsurprisingly, all methods are quite accurate at identifying the dominant eigenvalue, though the variants proposed in the present work show improvements in both the mean and scatter over the 1000 trials. In terms of the hidden eigenvalue, we observe that DMD (as well as OMD) estimates a decay rate that is almost double the true value. In contrast, all of ncDMD, fbDMD, and tlsDMD predict the eigenvalue accurately, with a reduction in the error of the mean eigenvalue between DMD and fbDMD (for example) of 88\%. In addition, we note that the scatter in the identified hidden eigenvalue across the trials is  smaller for fbDMD and tlsDMD (as indicated by smaller confidence ellipses). 

\subsection{Differentiating between process and sensor noise}
\label{sec:process}
\begin{figure}
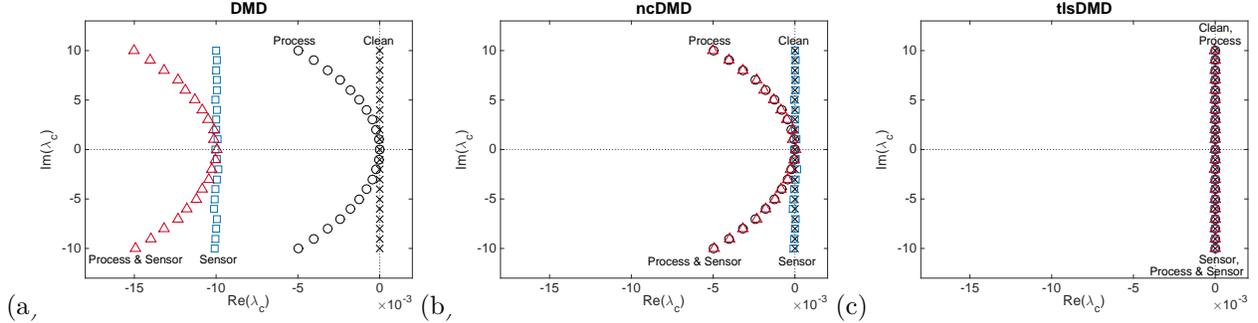

  (a)\includegraphics[width= 0.3\textwidth]{figSLa.eps}
  (b)\includegraphics[width= 0.3\textwidth]{figSLb.eps}
  (c) \includegraphics[width= 0.3\textwidth]{figSLc.eps}
\caption{Eigenvalues identified using (a) DMD, (b) ncDMD, and (c) tlsDMD for the Stuart-Landau equation (Eq.~\eqref{eq:SL}), with $100,000$ snapshots of data from Eq.~\eqref{eq:SLdata}, with $r_0 = 1$, $\mu = 1$, $\gamma = 1$, $\beta = 0$, and $dt  = 0.01$. Data with sensor noise, process noise, neither and both are considered, with noise levels for process and sensor noise being $\sigma_P^2 = 0.01$ and $\sigma_N^2 = 10^{-4}$ respectively. Note that in the absence of sensor noise, DMD and ncDMD are identical}
\label{fig:SLeigs}      
\end{figure}

This section will primarily address the issue of comparing and distinguishing between the effects of process and sensor noise. 
%
%
We consider the Stuart-Landau equation, which has been used as a model for the transient and periodic dynamics of flow past a cylinder in the vortex shedding regime \citep{noack:03cyl,Bagheri2013koopman}. 
In discrete time, we can express this system in polar coordinates by
\begin{equation}
\begin{aligned}
\label{eq:SL}
r_{k+1} &= r_k + dt(\mu r_k-r_k^3+n_r), \\
\theta_{k+1} &= \theta_k+dt(\gamma-\beta r_k^2+\frac{n_\theta}{r_k}),
\end{aligned}
\end{equation}
where we have included process noise terms $n_r$ and $n_\theta$, which are assumed to be independent in time, and sampled from separate zero-mean Gaussian distributions with variance $\sigma_P^2$. 
We take as our data snapshots of the form 
\begin{equation}
\label{eq:SLdata}
\bx_k = \begin{bmatrix} e^{-J i \theta_k} \ e^{(-J+1) i \theta_k} \  \cdots \ e^{J i \theta_k} \end{bmatrix}^T,
\end{equation}
for some integer $J$. We may add sensor noise to this data as in previous sections. For $\mu > 0$, Eq.~\eqref{eq:SL} contains a stable limit cycle at $r = \sqrt{\mu}$, with period $2\pi/(\gamma-\beta\mu)$. 
Starting on the limit cycle, we consider data with process noise, sensor noise, neither, and both. Without any noise, the eigenvalues identified from this data will lie upon the imaginary axis, at locations given by $\lambda_c = i j (\gamma - \mu \beta)$. 
Process noise acts to perturb the system from its limit cycle, which ultimately leads to phase diffusion, and a ``bending" of the eigenvalues such that they instead lie on a parabola. 
The behavior of this system with process noise is described more extensively in \cite{bagheri2014noise}. 
Fig.~\ref{fig:SLeigs} shows the results of applying variants of DMD on data generated by Eq.~\eqref{eq:SL} with  $\mu = 1$, $\gamma = 1$, $\beta = 0$, and $dt  = 0.01$, with data collected using  Eq.~\eqref{eq:SLdata} with $J =10$. 
Applying DMD on noise-free data gives eigenvalues along the imaginary axis, while data from the system with process noise gives a parabola of eigenvalues, as expected. 
For data collected using Eq.~\eqref{eq:SLdata}, each data channel will be orthogonal in time, and will contain the same energy. As a result, sensor noise will act to shift all identified eigenvalues into the left half plane by the same amount, as observed in Fig.~\ref{fig:SLeigs}(a). 
Fig.~\ref{fig:SLeigs}(b) shows that applying ncDMD accurately corrects for this shift, for the system with and without process noise. 
This shows that it is possible to distinguish between the effects of these two forms of noise, given only an estimate of the magnitude of the sensor noise. 
That is, we are able to eliminate the effects of the noise that is due to imperfections in our observations, while retaining the effects of actual disturbances to the system. 
Fig.~\ref{fig:SLeigs}(c) shows the tlsDMD corrects for the effects of both process and sensor noise, which is desirable if one wishes to recover the dynamics of the noise-free system. 
The results for fbDMD are not shown, but were very similar to those for tlsDMD. 
The ability of tlsDMD and fbDMD on process noise is not surprising, since they treat $X$ and $Y$ in a symmetric manner, and thus consider phase diffusion both forwards and backwards in time.

\section{Results with numerical and experimental data}
\label{sec:3}

Having analyzed the performance of the various proposed modifications of DMD on synthetic data sets, we now turn our attention to data obtained from fluids simulations and experiments. We will focus on the canonical case of the unsteady wake of a circular cylinder exhibiting periodic vortex shedding. In Sect.~\ref{sec:DNScyl} we present results from data obtained from a two-dimensional direct numerical simulation, while Sect.~\ref{sec:Expcyl} considers data obtained from PIV measurements in a water channel.

\subsection{Cylinder wake: simulation data}
\label{sec:DNScyl}
We use an immersed boundary projection method \citep{taira:07ibfs,taira:fastIBPM}, with a domain consisting of a series of nested grids, with the finest grid enclosing the body, and each successive grid twice as large as the previous. The finest grid consists of uniformly spaced points with grid spacing equal to $0.02D$ (where $D$ is the cylinder diameter), extending $2D$ upstream and $4D$ downstream of the center of the cylinder, and spanning $4D$ in the direction normal to the flow. Each successively larger grid contains the same number of grid points, with twice the grid spacing as the previous grid. The coarsest grid spans $24D$ in the streamwise direction and $16D$ in the normal direction. Uniform boundary conditions are used to first solve the Navier-Stokes equations on the largest grid, with each smaller grid using the next larger grid for boundary conditions. 
The numerical scheme uses a 3rd order Runge-Kutta time stepper, with a time step of $0.01 D/U_\infty$, where $U_\infty$ and $D$ are the freestream velocity and cylinder diameter, respectively.
The Reynolds number $\Rey  = \frac{U_\infty D}{\nu}$ was set to be 100, where $\nu$ is the kinematic velocity. This Reynolds number is above that for which the wake is stable (47, \cite{provansal1987benard}), and below that for which 3-dimensional instabilities emerge (approximately 194, \cite{williamson1996vortex}). At this Reynolds Number, the wake is hence unstable, and approaches a single periodic limit cycle characterized by a von-K\`arm\`an vortex street in the wake. The data to be analyzed was taken from 234 snapshots of the vorticity field, spaced $0.1D/U$ time units apart. This corresponds to approximately 4 complete periods of vortex shedding.
We truncate the data to only consider first 15 POD modes. These first 15 POD modes contain 99.99\% of the total energy of the clean data, and 92.96\% of the total energy of the data after the addition of Gaussian white noise with standard deviation $\sigma = 0.2 U/D$. Thus it is almost entirely noise that is truncated for the noisy data.

\begin{figure}
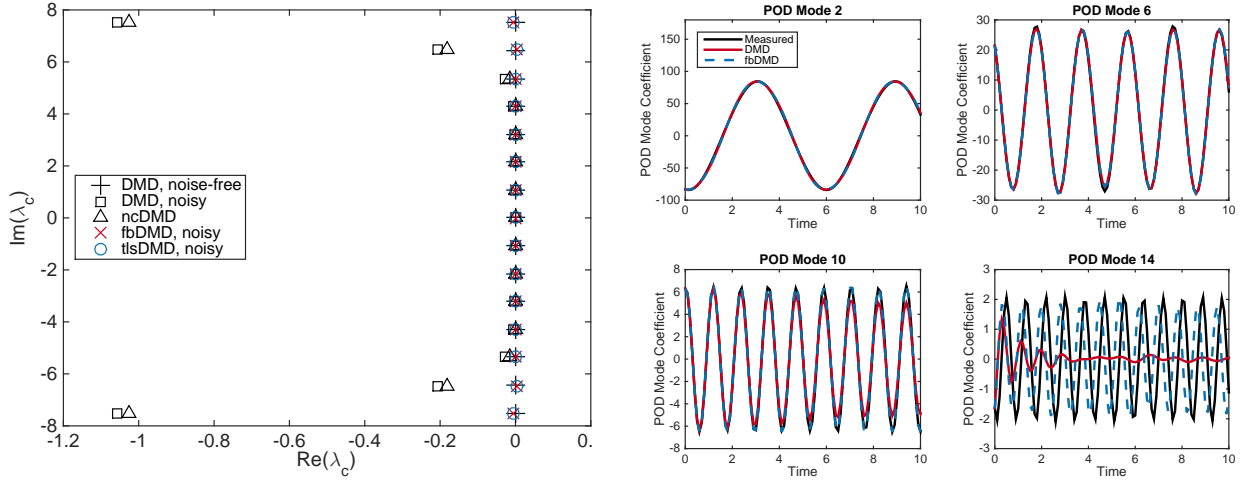

\center{(a)\includegraphics[width= 0.48\textwidth]{Fig9a.eps}}(b)\includegraphics[width= 0.48\textwidth]{Fig9b.eps}
\caption{(a) Eigenvalues and (b) POD coefficients identified from applying DMD, ncDMD fbDMD, and tlsDMD to DNS vorticity data from a cylinder wake at $Re = 100$. Noisy data was corrupted with Gaussian white noise with $\sigma = 0.2 U/D$ }
\label{fig:CylinderDNSeigs} 
\end{figure}

Fig.~\ref{fig:CylinderDNSeigs} shows results from applying various variants of DMD to such data. Though not shown, the results of applying tlsDMD were visually indistinguishable as using fbDMD. Since we are artificially adding noise, we can compare the results using noisy data to those generated from the noise-free data. When applying regular DMD to noisy data, we observe significant errors in the growth rate associated with the highest-frequency eigenvalues (Fig.~\ref{fig:CylinderDNSeigs}(a)). For an oscillatory system such as this, the DMD eigenmodes are very similar to the POD modes, with a DMD mode corresponding to $\lambda_c \approx 0$ that is almost the mean flow, and the modes associated with conjugate pairs of DMD eigenvalues corresponding to pairs of POD modes with equal energy, see \cite{chen2011variants} for further discussion of this phenomenon. This means that the observed measured eigenvalues are in line with the analysis given in Sect.~\ref{sec:DMDnoise} and Appendix 1, since the lower energy POD modes oscillate the most. We can see the effect of this error in Fig.~\ref{fig:CylinderDNSeigs}(b), which shows the prediction of a number of POD coefficients as evolved by the identified system, starting from the true initial condition. The dominant, low frequency POD modes are accurately predicted, but the higher ``harmonics'' are erroneously predicted to decay when using regular DMD. ncDMD improves the performance marginally, while fbDMD and tlsDMD both almost completely remove the erroneous decay of the high-frequency modes.

\begin{figure}
(a)\includegraphics[width= 0.63\textwidth]{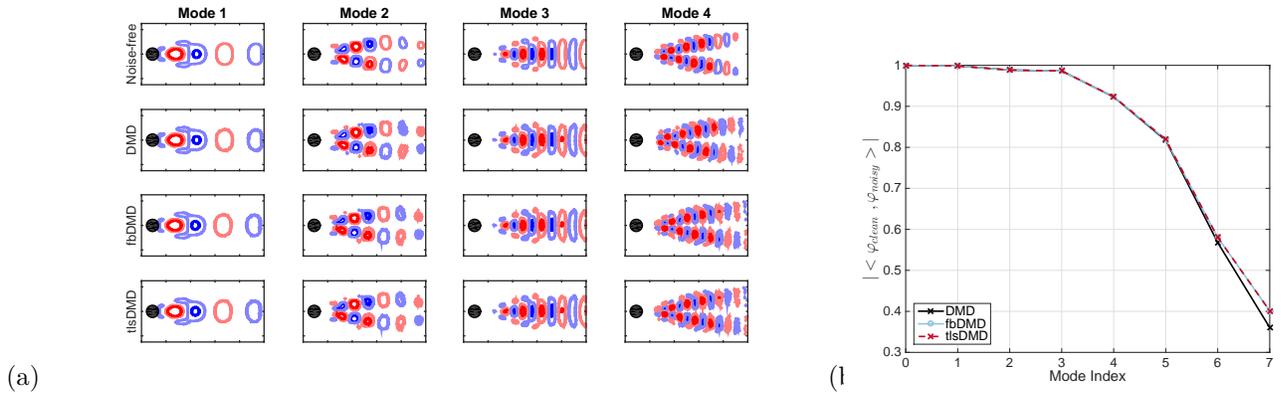}
(b)\includegraphics[width= 0.33\textwidth]{Fig10b}
\caption{(a) A subset of the DMD modes (real components) computed from applying various variants of DMD to DNS data of flow around a cylinder. (b) Inner product between (normalized) clean modes, and modes obtained from noisy data (with $\sigma_N = 0.2 U_\infty/D$) }
\label{fig:CylinderDNSmodes}      
\end{figure}

As well as considering eigenvalues, we also validate in
Fig.~\ref{fig:CylinderDNSmodes} that the modifications of DMD do not adversely
affect the identified DMD modes. This is shown both visually in
Fig.~\ref{fig:CylinderDNSmodes}(a), and quantitatively in
Fig.~\ref{fig:CylinderDNSmodes}(b), where we give the inner product
$\ip< \phi_{i,noisy}, \phi_{i,clean}>$ of the $i^{th}$ modes identified from
clean and noisy data, where we have pre-scaled the modes to be of unit norm. We
enumerate the modes by the imaginary component of the associated eigenvalue,
with mode 0 corresponding to the eigenvalue on the real axis. 
For modes that come in complex conjugate pairs, we arbitrarily consider those with positive
imaginary component.
We see that both fbDMD and tlsDMD marginally
outperform regular DMD, in terms of identifying modes that are at least as close
to those identified from noise-free data. The decrease in the inner product as
the mode number increases is indicative of noise being more significant in
higher-frequency modes, which contain less energy.

\subsection{Cylinder wake: experimental data}
\label{sec:Expcyl}
\begin{figure}
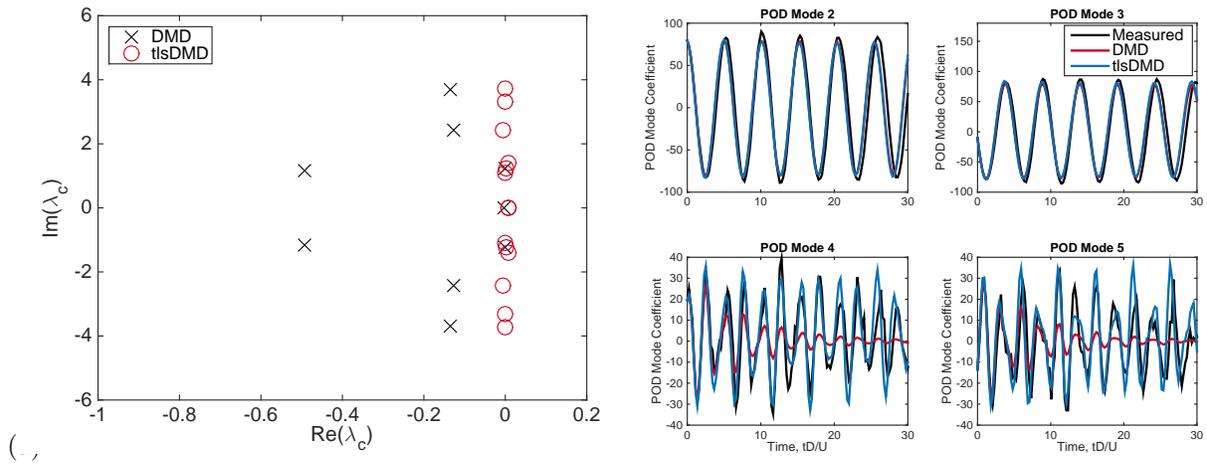

\center{(a)\includegraphics[width= 0.45\textwidth]{Fig11a.eps} 
(b)\includegraphics[width= 0.45\textwidth]{Fig11b.eps}}
\caption{(a) Eigenvalues and (b) POD coefficients identified from applying DMD and tlsDMD to experimental vorticity data}
\label{fig:CylinderExp}      
\end{figure}

We now turn our attention to data acquired from water channel experiment. An anodized aluminum cylinder of diameter $D = 9.5 \ mm$ and length $L = 260 \ mm$ was immersed in a recirculating, free-surface water channel with freestream velocity $U_\infty = 4.35 cm/s$. This gives a Reynolds number $\Rey = \frac{D U_\infty}{\nu} = 413$.  
Further details of the experimental setup and methodology are provided in \cite{tu2014compressed}. We apply variants of DMD to 500 snapshots from a vorticity field of size $135 \times 80$. Fig.~\ref{fig:CylinderExp} shows the identified eigenvalues and the predicted POD coefficients from the models identified from DMD and tlsDMD.  As in Sect.~\ref{sec:DNScyl}, we first project onto the 15 most energetic POD modes. Note that some eigenvalues (typically quickly decaying) are outside the range of the plot. As was the case with DNS data, we observe that DMD gives eigenvalues that are further into the left half plane that and of the other methods. This manifests in the erroneous prediction of decaying POD coefficients (Fig.~\ref{fig:CylinderExp}(b)), particularly for modes that are less energetic, and more rapidly oscillating. We thus conclude that more accurate low-dimensional models for the experimental results can be achieved by using tlsDMD. 
 We note that this improvement can be attained without explicit knowledge of the process and sensor noise characteristics. 

\section{Discussion and conclusions}
It was shown in Sect.~\ref{sec:2} that simple linear algebraic considerations can allow us to derive an estimate for the bias that exists in all standard formulations of DMD.  This subsequently led to the formulation of the three modified algorithms that we suggest can be used to eliminate this bias. Sect.~\ref{sec:resultssynth}  showed that this predicted bias is indeed present in the results of DMD. 
Directly correcting for this bias term (Algorithm \ref{alg:2}, ncDMD) was shown to almost completely eliminate this bias. While this modification demonstrates that our characterization of the dominant effects of noise was accurate, its usefulness is limited by the fact that it requires an accurate estimate of the noise covariance. Additionally, the presence of a $\Sigma^{-2}$ term in correction factor used in ncDMD makes this computation unsuitable for cases where small singular values that are not truncated. On the other hand, the correction factor in Algorithm \ref{alg:2} may be applied to existing DMD results with minimal computational effort.  Algorithms \ref{alg:3} (fbDMD) and \ref{alg:4} (tlsDMD), which do not require knowledge of the noise characteristics, were also found to correct for the bias, and also were able to reduce the random error across many noise realizations (as seen by smaller associated confidence ellipses in Fig.~\ref{fig:eigs1}, for example). 
Furthermore, fbDMD and tlsDMD were found in Sect.~\ref{sec:process} to also compensate for the effect of process noise. This feature could be desirable or undesirable, depending on the purpose for which DMD is being applied. 
Note that this is also consistent for the findings in Sect.~\ref{sec:Expcyl}, where for a notionally periodic system, tlsDMD was found to give eigenvalues very close to the imaginary axis, despite (presumably) the presence of both sensor and process noise.

In practice, the examples examined in Sects.~\ref{sec:hidden}, \ref{sec:DNScyl} and \ref{sec:Expcyl} suggest an overarching principle: while regular DMD can be accurate for identifying dominant dynamics that have much larger amplitudes than the noise in the data, accurate identification of the eigenvalues associated with lower amplitude modes (and in particular, their real components) can be significantly improved when using the modified DMD algorithms presented here. Conversely, if one is primarily concerned with the identification of modes and their frequencies of oscillation, and less concerned with accurate identification of growth/decay rates, then the effect of sensor noise is comparatively minimal, and subsequently the choice of DMD algorithm is less important.

Fundamentally, the bias in DMD arises because the algorithm is essentially a least-squares algorithm, which is designed for cases where the ``independent'' variable (which in DMD takes the form of the data $X$) is known accurately, and the ``dependent'' variable~$Y$ contains the noise/error. In reality, since $X$ and~$Y$ should both be affected by noise, minimizing the error in both the $X$ and~$Y$ ``directions'' can allow for a more accurate answer to be obtained. One drawback of tlsDMD is that it requires taking the SVD of a larger matrix. 
Note that for cases where $n > m$ (i.e., the size of each snapshot is larger than the number of snapshots) and there is no truncation of POD modes corresponding to small singular values, DMD gives the minimum Frobenius norm (of $A$) solution to $AX = Y$. In this case, in principle neither fbDMD or tlsDMD should yield any improvements. In reality, however, if there is noise in the data, then we do not necessarily want an exact fit to the data, but rather an unbiased estimate of the noise-free dynamics. We may obtain this by truncating POD modes that are deemed to be mostly noise, and use some variant of DMD to identify the remaining dynamics.
tlsDMD and fbDMD give very similar results, which suggests that fbDMD can be viewed as a computationally cheaper alternative to approximating the results of tlsDMD. Note that while fbDMD is often computationally cheaper, it relies on being able to invert the matrix $\tilde B_m$, which might be an ill-conditioned operation for some data.

In Sect.~\ref{sec:MethodsCompare}, we compared the variants presented here with two recent optimization algorithms that have been proposed. The results show our algorithms outperforming both sparsity promoting DMD and OMD. Note that since these algorithms are not in closed form, but instead contain optimization procedures, the results depend somewhat on the specification of the relevant optimization parameters. 
 In this comparison, our use of Algorithms \ref{alg:2}--\ref{alg:4} relied upon the projection onto a low dimensional subspace before applying DMD-type algorithms. 
We particularly note that the tlsDMD algorithm proposed here is slightly different from that given in \cite{hemati2015tls} due to this POD projection, which we found empirically to give improved results. We suggest that this is because the initial truncation of low-energy POD modes has a filtering effect that better isolates the true dynamics, at least for the datasets considered here.
 One could imagine, however, that in certain cases this projection could lead to significant degradation of results.
For example, where the dynamically important modes are highly dissimilar to the dominant POD modes, the flexibility for the projection basis to be modified could be particularly advantageous. In such cases, sparsity promoting DMD or OMD could give more favorable results.
In general, it is relatively common in system identification to use a subspace that is larger than the dimension of the underlying dynamics, and then later truncate to obtain a reduced order model of an appropriate size/rank. This can be particularly important when the dealing with specific system inputs and outputs \citep{rowley:05pod}. \cite{juang1986noise} discuss a number of ways in which true dynamic modes can be distinguished for noisy modes, in the context of the eigensystem realization algorithm. \cite{tu2014dynamic} further discusses how DMD modes can be scaled, from which appropriate modes can be chosen. The spDMD algorithm in~\cite{jovanovic2014dmdsp} essentially automates this procedure, and comes with the additional potential advantage of not requiring a-priori knowledge of the dimension of the reduced order dynamics to be identified. Note that it is also possible to combine the modifications to DMD proposed here with the OMD and spDMD optimization procedures, which could result in further improvements in some circumstances.

Though we used a large number of trials when testing our results on synthetic data in order to obtain statistically meaningful findings, in reality one would most likely not have this luxury with real data. In this case, it is important to understand for the size and quality of the data to be analyzed, both the best algorithm to use, and the amount of confidence that should be had in the results of the chosen algorithm.

While this work has been motivated by and has largely focused on sensor noise (that is, noise which only affects measurements, and not the system dynamics), the characterization and removal of process noise (i.e., disturbances to the system states) is entirely another matter.  Interestingly, the effect of process noise was identified analytically in \cite{bagheri2014noise} to be a parabolic decay in the growth rate of identified eigenvalues with increasing frequency. It turns out that a similar effect is observed here for the case of measurement noise. 
Isolating sensor noise from process noise (especially with limited prior information about the statistics of either) is an important and challenging task, particularly when dealing with more complex, turbulent flows, where the true dynamics exist on a wide range of spatial and temporal scales. The fact that DMD, ncDMD and tlsDMD/fbDMD each perform differently on these different forms of noise could itself be an important tool to this end.

Particularly in experimental data, users might typically preprocess data in a number of ways before considering applying DMD-type algorithms. It could be advantageous to investigate precisely how various averaging and smoothing operations affect the subsequent analysis of dynamics, and subsequently whether such post-processing and analysis can be achieved through a single algorithm. 

Ultimately, having a larger selection of possible algorithms should be of benefit to researchers who desire the dynamical information that DMD-type algorithms can provide, who can choose based on the size of the data, amount of noise present, required accuracy of the results, and amount of computational resources available.
One of the major advantages of DMD (and related algorithms) advocated in \cite{schmid2010DMD} is the fact that it requires only direct data measurements, without needing knowledge of any underlying system matrix, thus making it well suited to use on experimentally acquired data. Inevitably, however, data from experiments is always affected to some extent by noise. It is thus important to properly understand and quantify how noise can influence the results of DMD. 
Conversely, the quest for high quality data can often require large investments of both time and money. Formulating algorithms that are more robust to noisy data can be a cheaper alternative to obtain results of sufficient accuracy.
As it becomes easier to generate and store increasingly large datasets, it is also important to recognize that simply feeding larger quantities of data (e.g., more snapshots) into a given algorithm does not guarantee desired improvements in the accuracy of their outputs, as illustrated in Figures \ref{fig:froberrors} and~\ref{fig:froberrorsmore}.

The problem that fluid dynamicists face in extracting tractable information from large datasets is not unique to fluids, and rather transcends a wide variety of fields of study (although other fields are often not afforded the luxury of knowing the underlying differential equations). It is valuable to recognize and make use of the parallels in previous and current developments across a wide range of fields. We likewise hope that other areas can benefit from the work that is motivated by the desire to understand how fluids flow.

\section*{Acknowledgements}
The authors gratefully acknowledge the support for this work from the Air Force Office of Scientific Research grant FA9550-14-1-0289, J. Shang for making available the PIV data for flow past a cylinder, and J. Tu for insightful discussions. 

\section*{Appendix 1: Quantifying the size of the bias in DMD}
 We seek to quantify the magnitude of this bias present in DMD that was derived in Sect.~\ref{sec:DMDnoise}, subject to certain simplifying assumptions on the nature of the data and noise. If the noise is uniform, and spatially and temporally independent, then $\bE( \tilde N_X\tilde N_X^* ) = \bE(U^*N_X N_X^*U) = U^*m \sigma_N^2U  = m \sigma_N^2 I$, where $\sigma_N^2$ is the variance of each independent component of the noise matrix. 
 Furthermore, if we assure that we are projecting onto the POD modes of the clean data, then $(\tilde X \tilde X^*)  = \Sigma^2$, where $ U \Sigma V^*$ is the singular value decomposition of $X$. Thus with these assumptions, Eq.~\eqref{eq:DMDev} can be simplified to give
\begin{equation}
\bE(\tilde A_m) = \tilde A(I - m\sigma_N^2\Sigma^{-2}).
\end{equation}
The (diagonal) entries $\Sigma_i^2$ of $\Sigma^2$ have the interpretation of being the energy content of the $i^{th}$ POD mode. 
We then should expect that $\Sigma_i^2 \sim mnq_i \sigma_X^2$, where $\sigma_X^2$ is the RMS value of the elements in the data matrix $X$, 
 and $q_i = \frac{\Sigma_i^2}{\text{Trace}(\Sigma^2)}$ is the proportion of the total energy of the system contained in the $i^{th}$ POD mode. 
For this scaling, we make the assumption that adding/removing rows and columns of data (i.e., varying $m$ and $n$) does not affect either $\sigma_X$ or $q_i$.
 The bias term $m\sigma_N^2\Sigma^{-2}$ is a diagonal matrix whose $i^{th}$ entry has a size $(e_b)_i$ proportional to 
\begin{equation}
\label{eq:biasscale}
(e_b)_i \sim \frac{1}{n q_i SNR^2},
\end{equation}
where $SNR$ is the signal-to-noise ratio.
Thus sensor noise has the effect of reducing the diagonal entries of the computed $\tilde{A}_m$ matrix by a multiplicative factor of $1-\frac{\sigma_N^2}{n q_i \sigma_X^2}$, which means that POD coefficients are predicted to decay more rapidly than they actually do.  This effect is most pronounced for lower energy modes, for which the $q_i$ is smaller. We thus expect to identify with DMD (continuous-time) eigenvalues that are further into the left half plane than they should be (or would be if we applied DMD to noise-free data). \cite{duke2012error} argues in the case of periodic data that the growth rate of the eigenvalues should typically be the most challenging to identify, since there are a range of pre-existing methods that can identify frequencies. Here we have argued that it is precisely this growth rate that is most affected by noise.
Importantly, the amount of bias is independent of $m$, which suggests that the bias component of the error will be particularly dominant when we have a large number of low-dimensional snapshots. Importantly, this suggests that one cannot always effectively reduce the effect of noise by simply using more snapshots of data, since the bias error will eventually become the dominant error. 

While we can now quantify the magnitude of the bias in DMD, we do not as yet know how it compares to the random component of the error that would arise from a given realization of noise. 
To do this, we will estimate the typical size of the variance of individual entries of $\tilde A$, using the standard definition
\begin{equation}
var\left[\tilde A_{ij} \right] = \bE\left\{ \left((\tilde Y_m \tilde X_m^+)_{ij} - \bE\left[ (\tilde Y_m \tilde X_m^+)_{ij}\right]\right)\left((\tilde Y_m\tilde X_m^+)_{ij} - \bE\left[ (\tilde Y_m \tilde X_m^+)_{ij}\right]\right)\right\}.
\end{equation}
Referring back to Eq.~\eqref{eq:DMDnoisefull}, if we exclude terms that are quadratic or higher in noise, and assume that the noise covariance matrix is sufficiently close to its expected value, we find that 
\begin{align*}
(\tilde Y_m \tilde X_m^+) - \bE\left[ (\tilde Y_m \tilde X_m^+)\right]
				&= (\tilde Y + \tilde N_Y)(\tilde X + \tilde N_X)(\tilde X \tilde X^* + \tilde X\tilde N_X^* + \tilde N_X \tilde X^*+\tilde N_X\tilde N_X^*)^{-1}
				      - \tilde Y\tilde X^+ - \bE(\tilde N_X\tilde N_X^*)\Sigma^{-2}\\
				&= \left[\tilde Y\tilde X^+(\tilde X \tilde N_X^* + \tilde N_X\tilde X^*)+ \tilde N_Y\tilde X^* + \tilde Y\tilde N_X^*\right]\Sigma^{-2}.
\end{align*}
Elements of the terms $\tilde X \tilde N_X^*,\  \tilde N_X\tilde Y^*, \ \tilde N_Y\tilde X^* $, and $\tilde Y\tilde N_X^*$ are uncorrelated sums over $m$ random terms, with each term in the sum having variance $n q_i\sigma_X^2\sigma_N^2$ where as before $q_i$ is the energy fraction in the $i^{\text th}$ POD mode. This means that the sum will have variance $mn q_i\sigma_X^2\sigma_N^2$. Assuming that $\tilde Y\tilde X^+ (=\tilde A)$ does not greatly change the magnitude of quantities that it multiplies, and assuming that $q_i$ remains constant when varying $m$ and $n$, this means that we find that 
\begin{equation}
var\left[\tilde A_{ij} \right] \sim \frac{\sigma_N^2}{mn\sigma_X^2}.
\end{equation}
Thus the expected size of the random error in applying DMD to noisy data is 
\begin{equation}
\label{eq:randomerror}
e_r \sim \frac{1}{m^{1/2}n^{1/2}SNR}.
\end{equation}
Comparing Eq.~\eqref{eq:randomerror} with Eq.~\eqref{eq:biasscale}, we propose that the bias in DMD will be the dominant source of error whenever 
\begin{align*}m^{1/2}SNR > n^{1/2}.\end{align*} 

\bibliographystyle{spbasic}      
\bibliography{DMDNoise}

\begin{thebibliography}{46}
\providecommand{\natexlab}[1]{#1}
\providecommand{\url}[1]{{#1}}
\providecommand{\urlprefix}{URL }
\expandafter\ifx\csname urlstyle\endcsname\relax
  \providecommand{\doi}[1]{DOI~\discretionary{}{}{}#1}\else
  \providecommand{\doi}{DOI~\discretionary{}{}{}\begingroup
  \urlstyle{rm}\Url}\fi
\providecommand{\eprint}[2][]{\url{#2}}

\bibitem[{Bagheri(2013)}]{Bagheri2013koopman}
Bagheri S (2013) {Koopman}-mode decomposition of the cylinder wake. Journal of
  Fluid Mechanics 726:596--623

\bibitem[{Bagheri(2014)}]{bagheri2014noise}
Bagheri S (2014) Effects of weak noise on oscillating flows: linking quality
  factor, {Floquet} modes, and {Koopman} spectrum. Physics of Fluids
  (1994-present) 26(9):--

\bibitem[{Bai and Silverstein(2009)}]{bai2009spectral}
Bai Z, Silverstein JW (2009) Spectral analysis of large dimensional random
  matrices. Springer

\bibitem[{Belson et~al.(2013)Belson, Tu, and Rowley}]{belson2013modred}
Belson BA, Tu JH, Rowley CW (2013) A parallelized model reduction library. ACM
  T Math Software

\bibitem[{Chen et~al.(2011)Chen, Tu, and Rowley}]{chen2011variants}
Chen KK, Tu JH, Rowley CW (2011) Variants of dynamic mode decomposition:
  boundary condition, {Koopman}, and {Fourier} analyses. Journal of Nonlinear
  Science 22(6):887--915

\bibitem[{Cheng and Singer(2013)}]{cheng2013spectrum}
Cheng X, Singer A (2013) The spectrum of random inner-product kernel matrices.
  Random Matrices: Theory and Applications 2(04)

\bibitem[{Colonius and Taira(2008)}]{taira:fastIBPM}
Colonius T, Taira K (2008) A fast immersed boundary method using a nullspace
  approach and multi-domain far-field boundary conditions. Computer Methods in
  Applied Mechanics and Engineering 197:2131--2146

\bibitem[{Duke et~al.(2012)Duke, Soria, and Honnery}]{duke2012error}
Duke D, Soria J, Honnery D (2012) An error analysis of the dynamic mode
  decomposition. Experiments in Fluids 52(2):529--542

\bibitem[{Eckart and Young(1936)}]{eckart1936approx}
Eckart C, Young G (1936) The approximation of one matrix by another of lower
  rank. Psychometrika 1(3):211--218

\bibitem[{Epps and Techet(2010)}]{epps2010error}
Epps BP, Techet AH (2010) An error threshold criterion for singular value
  decomposition modes extracted from {PIV} data. Experiments in fluids
  48(2):355--367

\bibitem[{Golub and Van~Loan(2012)}]{golub2012matrix}
Golub GH, Van~Loan CF (2012) Matrix computations, vol~3. JHU Press

\bibitem[{Goulart et~al.(2012)Goulart, Wynn, and Pearson}]{goulart2012omd}
Goulart PJ, Wynn A, Pearson D (2012) Optimal mode decomposition for high
  dimensional systems. In: CDC, pp 4965--4970

\bibitem[{Grosek and Kutz(2014)}]{grosek2014dmd}
Grosek J, Kutz JN (2014) Dynamic mode decomposition for real-time
  background/foreground separation in video. arXiv preprint arXiv:14047592

\bibitem[{Hemati et~al.(2014)Hemati, Williams, and
  Rowley}]{Hemati2014streaming}
Hemati MS, Williams MO, Rowley CW (2014) Dynamic mode decomposition for large
  and streaming datasets. Physics of Fluids (1994-present) 26(11):111,701

\bibitem[{Hemati et~al.(2015)Hemati, Rowley, Deem, and
  Cattafesta}]{hemati2015tls}
Hemati MS, Rowley CW, Deem EA, Cattafesta LN (2015) De-biasing the dynamic mode
  decomposition for applied koopman spectral analysis. arXiv preprint
  arXiv:150203854

\bibitem[{Ho and Kalman(1965)}]{Kalman:1965}
Ho BL, Kalman RE (1965) Effective construction of linear state-variable models
  from input/output data. In: Proceedings of the 3rd Annual Allerton Conference
  on Circuit and System Theory, pp 449--459

\bibitem[{Jardin and Bury(2012)}]{jardin2012dmd}
Jardin T, Bury Y (2012) Lagrangian and spectral analysis of the forced flow
  past a circular cylinder using pulsed tangential jets. Journal of Fluid
  Mechanics 696:285--300

\bibitem[{Jovanovi{\'c} et~al.(2014)Jovanovi{\'c}, Schmid, and
  Nichols}]{jovanovic2014dmdsp}
Jovanovi{\'c} MR, Schmid PJ, Nichols JW (2014) Sparsity-promoting dynamic mode
  decomposition. Physics of Fluids (1994-present) 26(2):--

\bibitem[{Juang and Pappa(1985)}]{ERA:1985}
Juang JN, Pappa RS (1985) An eigensystem realization algorithm for modal
  parameter identification and model reduction. Journal of Guidance, Control
  and Dynamics 8(5):620--627

\bibitem[{Juang and Pappa(1986)}]{juang1986noise}
Juang JN, Pappa RS (1986) Effects of noise on modal parameters identified by
  the eigensystem realization algorithm. Journal of Guidance, Control, and
  Dynamics 9(3):294--303

\bibitem[{Juang et~al.(1988)Juang, Cooper, and Wright}]{juang1987eradc}
Juang JN, Cooper J, Wright J (1988) An eigensystem realization algorithm using
  data correlations ({ERA/DC}) for modal parameter identification.
  Control-Theory and Advanced Technology 4(1):5--14

\bibitem[{Koopman(1931)}]{koopman1931hamiltonian}
Koopman BO (1931) Hamiltonian systems and transformation in hilbert space.
  Proceedings of the National Academy of Sciences of the United States of
  America 17(5):315

\bibitem[{Mezi{\'c}(2005)}]{mezic2005spectral}
Mezi{\'c} I (2005) Spectral properties of dynamical systems, model reduction
  and decompositions. Nonlinear Dynamics 41(1-3):309--325

\bibitem[{Mezi{\'c}(2013)}]{mezic2013koopman}
Mezi{\'c} I (2013) Analysis of fluid flows via spectral properties of the
  {Koopman} operator. Annual Review of Fluid Mechanics 45:357--378

\bibitem[{Noack et~al.(2003)Noack, Afanasiev, Morzynski, Tadmor, and
  Thiele}]{noack:03cyl}
Noack BR, Afanasiev K, Morzynski M, Tadmor G, Thiele F (2003) A hierarchy of
  low-dimensional models for the transient and post-transient cylinder wake.
  Journal of Fluid Mechanics 497:335--363

\bibitem[{Pan et~al.(2015)Pan, Xue, and Wang}]{pan2015accuracy}
Pan C, Xue D, Wang J (2015) On the accuracy of dynamic mode decomposition in
  estimating instability of wave packet. Experiments in Fluids 56(8):1--15

\bibitem[{Provansal et~al.(1987)Provansal, Mathis, and
  Boyer}]{provansal1987benard}
Provansal M, Mathis C, Boyer L (1987) {B{\'e}nard-von K{\'a}rm{\'a}n}
  instability: transient and forced regimes. Journal of Fluid Mechanics
  182:1--22

\bibitem[{Pyatykh et~al.(2013)Pyatykh, Hesser, and Zheng}]{Pyatykh2013image}
Pyatykh S, Hesser J, Zheng L (2013) Image noise level estimation by principal
  component analysis. Image Processing, IEEE Transactions on 22(2):687--699

\bibitem[{Rowley(2005)}]{rowley:05pod}
Rowley CW (2005) Model reduction for fluids using balanced proper orthogonal
  decomposition. International Journal of Bifurcation and Chaos 15(3):997--1013

\bibitem[{Rowley et~al.(2009)Rowley, Mezi{\'c}, Bagheri, Schlatter, and
  Henningson}]{rowley2009spectral}
Rowley CW, Mezi{\'c} I, Bagheri S, Schlatter P, Henningson DS (2009) Spectral
  analysis of nonlinear flows. Journal of Fluid Mechanics 641(1):115--127

\bibitem[{Schmid et~al.(2011)Schmid, Li, Juniper, and Pust}]{schmid2011tcfd}
Schmid P, Li L, Juniper M, Pust O (2011) Applications of the dynamic mode
  decomposition. Theoretical and Computational Fluid Dynamics 25(1-4):249--259

\bibitem[{Schmid(2010)}]{schmid2010DMD}
Schmid PJ (2010) Dynamic mode decomposition of numerical and experimental data.
  Journal of Fluid Mechanics 656:5--28

\bibitem[{Schmid(2011)}]{schmid2011expfluids}
Schmid PJ (2011) Application of the dynamic mode decomposition to experimental
  data. Experiments in Fluids 50(4):1123--1130

\bibitem[{Schmid and Sesterhenn(2008)}]{schmid2008}
Schmid PJ, Sesterhenn J (2008) Dynamic mode decomposition of numerical and
  experimental data. In: 61st Annual Meeting of the APS Division of Fluid
  Dynamics, American Physical Society

\bibitem[{Singer and Wu(2013)}]{singer2013two}
Singer A, Wu HT (2013) Two-dimensional tomography from noisy projections taken
  at unknown random directions. SIAM Journal on Imaging Sciences 6(1):136

\bibitem[{Stewart(1991)}]{stewart1991perturbation}
Stewart G (1991) Perturbation theory for the singular value decomposition. In
  SVD and Signal Processing, 11: Algorithms, Analysis and Applications

\bibitem[{Stewart(2006)}]{stewart2006perturbation}
Stewart M (2006) Perturbation of the {SVD} in the presence of small singular
  values. Linear Algebra and its Applications 419(1):53--77

\bibitem[{Taira and Colonius(2007)}]{taira:07ibfs}
Taira K, Colonius T (2007) The immersed boundary method: a projection approach.
  Journal of Computational Physics 225(2):2118--2137

\bibitem[{Tao and Vu(2012)}]{tao2012random}
Tao T, Vu V (2012) Random covariance matrices: universality of local statistics
  of eigenvalues. The Annals of Probability 40(3):1285--1315

\bibitem[{Tu et~al.(2014{\natexlab{a}})Tu, Rowley, Kutz, and
  Shang}]{tu2014compressed}
Tu JH, Rowley CW, Kutz JN, Shang JK (2014{\natexlab{a}}) Spectral analysis of
  fluid flows using sub-{Nyquist}-rate {PIV} data. Experiments in Fluids
  55(9):1--13

\bibitem[{Tu et~al.(2014{\natexlab{b}})Tu, Rowley, Luchtenburg, Brunton, and
  Kutz}]{tu2014dynamic}
Tu JH, Rowley CW, Luchtenburg DM, Brunton SL, Kutz JN (2014{\natexlab{b}}) On
  dynamic mode decomposition: theory and applications. Journal of Computational
  Dynamics 1(2):391--421

\bibitem[{Williams et~al.(2014)Williams, Rowley, and
  Kevrekidis}]{williams2014kernel}
Williams MO, Rowley CW, Kevrekidis IG (2014) A kernel approach to data-driven
  {Koopman} spectral analysis. arXiv preprint arXiv:14112260

\bibitem[{Williams et~al.(2015)Williams, Kevrekidis, and
  Rowley}]{williams2014edmd}
Williams MO, Kevrekidis IG, Rowley CW (2015) A data-driven approximation of the
  {Koopman} operator: Extending dynamic mode decomposition. Journal of
  Nonlinear Science pp 1--40

\bibitem[{Williamson(1996)}]{williamson1996vortex}
Williamson CH (1996) Vortex dynamics in the cylinder wake. Annual review of
  fluid mechanics 28(1):477--539

\bibitem[{Wynn et~al.(2013)Wynn, Pearson, Ganapathisubramani, and
  Goulart}]{wynn2013omd}
Wynn A, Pearson DS, Ganapathisubramani B, Goulart PJ (2013) Optimal mode
  decomposition for unsteady flows. Journal of Fluid Mechanics 733:473--503

\bibitem[{Zhao and Singer(2013)}]{zhao2013fourier}
Zhao Z, Singer A (2013) {Fourier--Bessel} rotational invariant eigenimages.
  JOSA A 30(5):871--877

\end{thebibliography}
\end{document}